\documentclass[prd,twocolumn,superscriptaddress, nofootinbib, preprintnumbers]{revtex4-1}
\usepackage{amsmath}
\usepackage{graphicx}
\usepackage{xcolor}
\usepackage{hyperref}
\newcommand{\Tau}{\mathrm{T}}
\usepackage{float}
\usepackage{braket}
\usepackage{comment}

\newcommand*{\defeq}{\mathrel{\vcenter{\baselineskip0.5ex \lineskiplimit0pt
			\hbox{\scriptsize.}\hbox{\scriptsize.}}}%
	=}

\newcommand{\be}{\begin{equation}}
\newcommand{\ee}{\end{equation}}

\begin{document}
\title{Generalized Boltzmann hierarchy for massive neutrinos in cosmology}
\author{Caio Bastos de Senna Nascimento \\{\it{\small  Department of Physics \& Astronomy, Stony Brook University, Stony Brook, NY 11794}}}

\begin{abstract}
	Boltzmann solvers are an important tool for the computation of cosmological observables in the linear regime. In the presence of massive neutrinos, they involve solving the Boltzmann equation followed by an integration in momentum space to arrive at the desired fluid properties, a procedure which is known to be computationally slow. In this work we introduce the so-called generalized Boltzmann hierarchy (GBH) for massive neutrinos in cosmology, an alternative to the usual Boltzmann hierarchy, where the momentum dependence is integrated out leaving us with a two-parameter infinite set of ordinary differential equations. Along with the usual expansion in multipoles, there is now also an expansion in higher velocity weight integrals of the distribution function. Using a toy code, we show that the GBH produces the density contrast neutrino transfer function to a $\lesssim 0.5\%$ accuracy at both large and intermediate scales compared to the neutrino free-streaming scale, thus providing a proof-of-principle for the GBH. We comment on the implementation of the GBH in a state of the art Boltzmann solver. 
\end{abstract}

\maketitle

\section{Introduction}
\label{sec:intro}

Neutrino oscillation experiments have established that neutrinos are massive particles (at least two eigenstates), with a lower bound, in the sum of all neutrino masses, of $\sum m_{\nu} \geq 0.06$eV, $0.1$eV for normal and inverted hierarchies, respectively \cite{1708.01186, 1806.11051}. The large-scale structure of our Universe gives a sensitive probe of neutrino masses \cite{1803.11545, astro-ph/0603494}. This allows us to use cosmological data to constrain the sum of neutrino mass eigenstates: $\sum m_{\nu} \lesssim (0.1-0.3)$eV, e.g. \cite{1811.02578, 1807.06209}, depending on the choice of used datasets. Current and future large-scale structure surveys \cite{1907.08945, 1910.09273, 1906.10134, 1305.5422, 1508.04473, 1111.6398, 1907.10688} will be used to determine the mass scale of neutrinos \cite{1903.03689, 1808.05955}, but also to constraint beyond-$\Lambda$CDM scenarios \cite{1903.12016, 1903.04763}. It is then of paramount importance that cosmological observables, such as the matter power spectrum, can be computed to a subpercent level accuracy, in both linear and nonlinear scales.

The study of structure formation in the nonlinear regime relies on N-body simulations \cite{astro-ph/0411043, astro-ph/0505010}. On the other hand, the linear theory is much simpler, and there are publicly available codes, such as the code for anisotropies in the microwave background (CAMB) \cite{lewis2011camb} and the cosmic linear anisotropy solving system (CLASS) \cite{1104.2932}, that can be used to compute the observables. The implementation of neutrinos in the linear theory is somewhat cumbersome, since it involves solving a Boltzmann hierarchy of equations in momentum space. The reason for this can be traced back to the usual statement that the momentum dependence in the distribution function cannot be integrated out \cite{astro-ph/9506072}. For this reason, fluid approximations have been developed in the past, and incorporated as an optional tool in the Boltzmann solvers \cite{1104.2935, astro-ph/9801234, astro-ph/0203507}.

In this work we show that the momentum dependence in the distribution function can, in fact, be exactly integrated out, at the expense of introducing a new countable parameter $n$, along with the parameter $l$ associated with the multipole expansion, to the infinite system of ordinary differential equations that need to be solved to determine the dynamics, in Fourier space (\cite{astro-ph/0607319, astro-ph/0203507} being examples of this in the literature). This leads us to a novel two-parameter infinite set of equations that determine the evolution of noncold dark matter (or ncdm, borrowing notation from CLASS \cite{1104.2935}) perturbations in a flat universe: The generalized Boltzmann hierarchy (GBH). Along with the usual multipole expansion, there is now also an expansion in higher velocity weight integrals of the distribution function \footnote{An expansion in higher velocity weights is used in CAMB to approximate the evolution of the perturbations in massive neutrinos once already in the non-relativistic regime \cite{astro-ph/0203507}, while in our approach we obtain the exact evolution, starting from initial conditions while still in the relativistic regime.}. The GBH is simpler than the usual approach, as implemented in Boltzmann solvers, in the sense that it does not require the numerical computation of momentum integrals, after solving the dynamical equations.    

The paper is organized as follows: In Sec. \ref{sec:gnfe}, we introduce the generalized Boltzmann hierarchy for ncdm perturbations, and compare it to the usual approach of evolving the distribution function in phase space.  In Sec. \ref{sec:nr}, we implement the equations numerically for a single massive neutrino component, with varying mass $m$ and scale $k$, obtaining $\lesssim 0.5 \%$ agreement with the Boltzmann solver CLASS in high precision settings, for all redshift. We also discuss the dependence of our framework on the neutrino mass $m$ and scale $k$, and show that when switching to a fluid approximation on the small scales, i.e. once a given mode becomes smaller than the free-streaming scale (hereafter named GBH+FA), we can produce the neutrino transfer function, at $z=0$, with the same accuracy as CLASS in its default precision settings (or CLASS-DPS), over all scales. We also found that CAMB, when also in its default precision settings (CAMB-DPS), yields more accurate results than both the GBH+FA and CLASS-DPS on the small scales. In Sec. \ref{sec:dc}, we conclude and comment on the implementation of the GBH in Boltzmann solvers, along with its current limitations. In Appendix \ref{sec:app}, we give a detailed account of the truncation scheme for the GBH. Finally, in Appendix \ref{sec:CLASS} we investigate the GBH in the simple case of $l_{\max}=2$ and $n_{\max}=0$, i.e. a viscous fluid approximation (FA). We compare the truncation scheme developed for the GBH with the one employed in CLASS.

\section{Generalized Boltzmann Hierarchy}
\label{sec:gnfe}

We start by introducing the additional expansion in higher velocity weight integrals of the distribution function, followed by a derivation of their associated dynamical equations. We will be considering scalar (linear) perturbations to a flat Friedmann-Robertson-Walker (FRW) universe, in the Newtonian gauge
\begin{equation}
	ds^2 = a(\tau)^2[-(1+2\psi)d\tau^2 +(1-2\phi)d\vec{x}^2]
\end{equation}
Conventions and notation follow \cite{astro-ph/9506072}. Let us first define suitable generalized fluid properties, at the level of background
\begin{equation}
\label{eq:bfp}
	P_{n} = \rho \omega_{n} \defeq \frac{4\pi}{3} a^{-4} \int_{0}^{\infty}  dq q^2 f_{0}(q) \epsilon \Big(\frac{q}{\epsilon}\Big)^{2n}  \ \forall n \geq 0
\end{equation}
where $f_{0}(q) = (2/(2\pi)^3) (1+e^{\frac{q}{T_{0}}})^{-1}$ is the Fermi-Dirac distribution written in terms of comoving momentum $\vec{q} = a\vec{p}$, with $\vec{p}$ the proper momentum and  $T_{0} \approx 1.95$K the temperature of relic neutrinos today. Also $\epsilon = \sqrt{q^2 + a^2 m^2} = aE$, with $m$ the neutrino mass and $E$ the proper energy. Then $P_{0} = (1/3) \rho$ is a third of the background energy density, $P_{1} = P$ is the pressure, $P_{2} \equiv \mathcal{P}$, and in general $P_{n+2} \equiv \mathcal{P}^{(n)}, n \geq 0$ are higher velocity weight pressures. Similarly, $\omega_{0} = 1/3$, $\omega_{1}  = \omega$ is the equation of state parameter, and $\omega_{n+2}, \ n \geq 0$ are higher velocity weight equation of state parameters. 

Taking the derivative of Eq.(\ref{eq:bfp}) with respect to conformal time gives the following hierarchy of equations
\begin{equation}
\label{eq:bfe}
\begin{split}
	\omega'_{n} = & -(2n+3)\mathcal{H} \omega_{n} + (2n-1)\mathcal{H}\omega_{n+1} \\ & -\frac{\rho'}{\rho} \omega_{n}  \  \forall n \geq 0
\end{split}
\end{equation}
where $\mathcal{H} = a'/a$, and $'$ denotes derivative with respect to conformal time. Notice that $q/\epsilon = v \sim T_{0}/ma$ is the physical velocity of an individual neutrino particle, such that the additional factors of $(q/\epsilon)^2$ in the integrals in Eq.(\ref{eq:bfp}) effectively shift the peak of the distribution function to higher particle velocities. In the relativistic regime, all particles travel at the speed of light, and one only needs to consider the $n=0$ equation. The same holds in the nonrelativistic regime, where $ \mathcal{O}(v^2)$ corrections become negligible. During the transition, however, the higher velocity weight fluid properties need to be taken into account, in order to probe the whole spectrum of neutrino particle velocities.  

Setting $n=0$ in Eq.(\ref{eq:bfe}) yields the familiar equation $\rho' + 3 \mathcal{H} (\rho +P)=0$. At this level, it is easier to simply determine the evolution of the distribution function, and then integrate Eq.(\ref{eq:bfp}) directly, than to approach the infinite set of Eqs. (\ref{eq:bfe}). This is because the background distribution function admits a simple, analytic solution (e.g. the Fermi-Dirac distribution). However, this is no longer true when inhomogeneities are introduced. 

In this case we have, along with the expansion in higher velocity weight integrals of the distribution function, parametrized by $n$, the usual expansion in multipoles, parametrized by $l$. To obtain it, split the distribution function as $f=f_{0}(q)(1+\Psi)$, and expand $\Psi$ in a Legendre series
\begin{equation}
\label{eq:multipoleexp}
	\Psi(\vec{k},\hat{n},q,\tau) = \sum_{l=0}^{\infty} (-i)^l(2l+1)\Psi_{l}(k,q,\tau)P_{l}(\hat{k} \cdot \hat{n})
\end{equation}
where we are working in Fourier space (set $\vec{\nabla} \to i\vec{k}$), and define $\hat{n} = \vec{q}/q$. 

In terms of the multipole expansion in Eq.(\ref{eq:multipoleexp}), the fluid properties directly sourcing the gravitational field, i.e. in the energy momentum tensor, are \cite{astro-ph/9506072}
\begin{subequations}
\label{eq:emt}
\begin{align}
		 \delta \rho = \rho \delta & = 4\pi a^{-4}  \int_{0}^{\infty} dq q^2 f_{0}(q)  \epsilon \Psi_{0} \\ \delta P & = \frac{4\pi}{3} a^{-4}  \int_{0}^{\infty} dq q^2 f_{0}(q)  \epsilon \Big(\frac{q}{\epsilon}\Big)^2 \Psi_{0} \\ (\rho +P)\theta & = 4\pi k a^{-4}  \int_{0}^{\infty}  dq q^2 f_{0}(q) q \Psi_{1} \\  (\rho +P)\sigma & = \frac{8\pi}{3} a^{-4}  \int_{0}^{\infty} dq q^2 f_{0}(q) \epsilon \Big(\frac{q}{\epsilon}\Big)^{2} \Psi_{2}  
\end{align}
\end{subequations}

\begin{figure}
		\centering
		\includegraphics[width=0.5\textwidth]{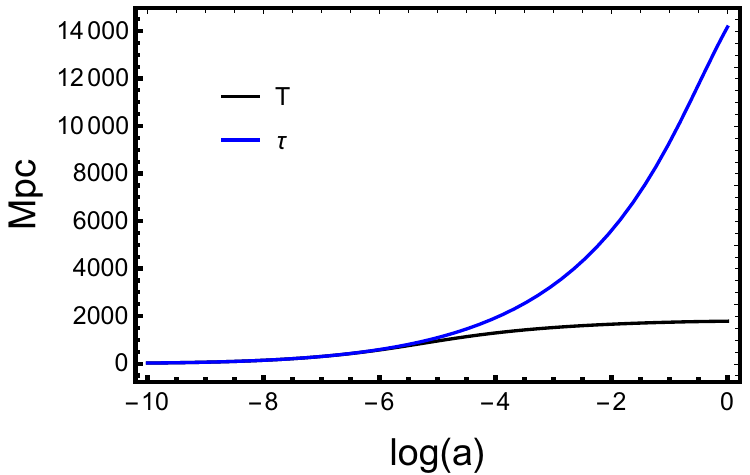}
		\caption{Evolution of the neutrino horizon, i.e. $\Tau$ for $q/T_{0} = 3$, and $m = 0.1$eV. It grows like the conformal time $\tau$ up to the time of transition, when it effectively freezes as it approaches the nonrelativistic regime. The significant difference between $\Tau$ and $\tau$ in the nonrelativistic regime explains why one needs a much higher $l_{\max}$ for radiation than for massive neutrinos \cite{astro-ph/9506072}. }
\label{fig:tau}
\end{figure}

\begin{figure}
		\centering
		\includegraphics[width=0.5\textwidth]{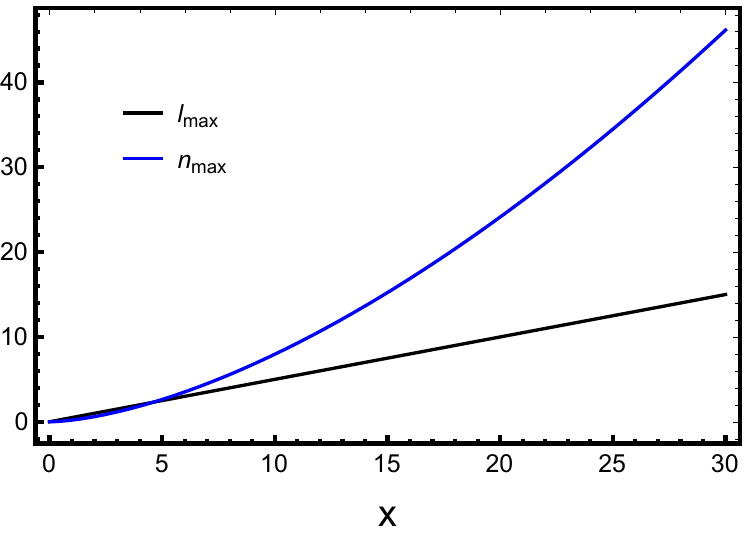}
		\caption{Choices of $l_{\max}$ and $n_{\max}$ as a function of $x$, as given by Eq.(\ref{eq:trunc}), to approximately ensure convergence, up to $x=30$. Note that $n_{\max}$ grows faster than linearly with $x$. }
\label{fig:lmaxnmax}
\end{figure}

We now wish to generalize this to higher multipoles, and also include higher velocity weight integrals of the distribution function, in analogy to Eq.(\ref{eq:bfp}), to make sure that the whole spectrum of particle velocities is being probed during the transition from the relativistic to nonrelativistic regimes. The following is then a natural choice of dynamical variables:
\begin{subequations}
\label{eq:fp}
\begin{align}
		 \delta P_{n} = \rho \delta_{n} & \defeq \frac{4\pi}{3} a^{-4}  \int_{0}^{\infty} dq q^2 f_{0}(q)  \epsilon \Big(\frac{q}{\epsilon}\Big)^{2n} \Psi_{0} \\  (\rho +P)\theta_{n} & \defeq 4\pi k a^{-4}  \int_{0}^{\infty}  dq q^2 f_{0}(q) \epsilon \Big(\frac{q}{\epsilon}\Big)^{2n+1} \Psi_{1} \\  (\rho +P)f_{l,n} & \defeq 4\pi \frac{l!}{(2l-1)!!} a^{-4}  \int_{0}^{\infty} dq q^2 f_{0}(q) \epsilon \Big(\frac{q}{\epsilon}\Big)^{2n+l} \Psi_{l} \\ \ \ \ \ \ \ \forall \  l \geq 1 \nonumber 
\end{align}
\end{subequations}
and $n \geq 0$ everywhere. Notice that $\delta_{0} = \frac{1}{3} \delta$ is a third of the density contrast, $\rho \delta_{1} = \delta P$ is the perturbation to the pressure, $\rho \delta_{2} \equiv \delta \mathcal{P}$, and $\rho \delta_{n+2} \equiv \delta \mathcal{P}^{(n)}, n \geq 0$ are the perturbations to the higher velocity weight pressures. Also $\theta_{0} = \theta$ is the divergence of the velocity, $\theta_{1} \equiv \Theta$ and $\theta_{n+1} \equiv \Theta^{(n)}, n \geq 0$ are its higher velocity weight counterparts. We also define $f_{2,n} \equiv \sigma_{n}$, the anisotropic shear stress, with a similar notation for its higher velocity weight integrals (i.e. $\Sigma^{(n)})$, and set $\theta_{n} \equiv kf_{1,n}$, when it is convenient to do so. In order to derive a set of equations for the variables in Eq.(\ref{eq:fp}), we need the time evolution of the multipoles. It follows from the substitution of Eq.(\ref{eq:multipoleexp}) into the Boltzmann equation, and reads \cite{astro-ph/9506072}
\begin{subequations}
\label{eq:eqm}
\begin{align}
		& \Psi'_{0} = -\frac{qk}{\epsilon} \Psi_{1} - \phi'\frac{d\ln f_{0}}{d\ln q} \\ & \Psi'_{1} = \frac{qk}{3\epsilon} (\Psi_{0}-2\Psi_{2}) - \frac{\epsilon k}{3q} \psi \frac{d\ln f_{0}}{d\ln q} \\ & \Psi'_{l} = \frac{qk}{(2l+1)\epsilon}[l\Psi_{l-1} - (l+1)\Psi_{l+1}]  \ \forall \ l \geq 2
\end{align}
\end{subequations}
Now take the derivative of each expression in Eq.(\ref{eq:fp}) with respect to conformal time, and use Eqs.(\ref{eq:bfp}), (\ref{eq:bfe}), (\ref{eq:fp}) and (\ref{eq:eqm}) to arrive at
\begin{subequations}
\label{eq:gfe}
\begin{align}
		& \delta_{n}' = -(2n-3w)\mathcal{H} \delta_{n} +(2n-1)\mathcal{H} \delta_{n+1} \\ & - \frac{1}{3} (1+w) \theta_{n} + [(2n+3)\omega_{n} - (2n-1)\omega_{n+1}]\phi' \nonumber \\ & \theta'_{n} = -\Big[(2n+1-3w)\mathcal{H} +\frac{w'}{1+w}\Big]\theta_{n} +2n\mathcal{H}\theta_{n+1} \\& + \frac{1}{1+w}k^2 \delta_{n+1}  -k^2 \sigma_{n} \nonumber \\ & + \frac{1}{1+w}[(2n+3)\omega_{n} - (2n-1)\omega_{n+1}]k^2 \psi \nonumber \\ & f'_{l,n} = -\Big[(2n+l-3w)\mathcal{H} + \frac{w'}{1+w}\Big]f_{l,n} \\ & +(2n+l-1)\mathcal{H}f_{l,n+1} + \frac{l^2}{4l^2-1}kf_{l-1,n+1} \nonumber \\ & -kf_{l+1,n} \ \forall l \geq 2  \nonumber
\end{align}
\end{subequations}
with $n \geq 0$. This is the generalized Boltzmann hierarchy (GBH) for a ncdm component. Setting $n=0$ in Eq.(\ref{eq:gfe}), one recovers the usual ncdm fluid equations (and up to $l=2$, including only dynamical equations for the fluid properties that directly source the gravitational field)
\begin{subequations}
\label{eq:ncdmfe}
\begin{align}
		& \delta' = -(1+w)(\theta-3\phi') - 3\mathcal{H}\Big(\frac{\delta P}{\delta \rho}-w\Big)\delta \\ & \theta' = -\Big[(1-3w)\mathcal{H} + \frac{w'}{1+w}\Big]\theta + \frac{\delta P/\delta \rho}{1+w} k^2 \delta - k^2 \sigma + k^2 \psi \\ & \sigma' = -\Big[(2-3w)\mathcal{H} + \frac{w'}{1+w}\Big]\sigma + \mathcal{H} \Sigma + \frac{4}{15} \Theta -kf_{3}
\end{align}
\end{subequations}
As is well known, Eq.(\ref{eq:ncdmfe}) involves variables, i.e. $\delta P/\delta \rho$, $\Sigma$, $\Theta$ and $kf_{3}$, that need to be somehow approximated, in terms of the dynamical variables in the system, in order to close the equations. On the other hand, the two-parameter hierarchy of Eqs.(\ref{eq:gfe}) is closed as it is, and we have achieved our goal: To get rid of the momentum integrals altogether. Of course, any practical implementation of the GBH requires a good truncation scheme: A choice of a given number of multipoles, $l_{\max}+1$, and higher velocity weight variables, $n_{\max}+1$, to dynamically evolve, together with a recipe for approximating higher order quantities. The resulting system of equations is of dimension $(l_{\max}+1) \times (n_{\max}+1)$. We discuss this at length in Appendix \ref{sec:app}. Here we will just spell out the recipe. The truncation in multipoles is done with the approximation
\begin{equation}
\label{eq:ltruncfp1}
	f_{l_{\max}+1,n} \approx (l_{\max}+1)\Big(\frac{1}{k\tau} f_{l_{\max},n} - \frac{l_{\max}}{4l_{\max}^2 -1} f_{l_{\max}-1,n+1}\Big)
\end{equation}
while the truncation in higher velocity weight integrals is handled with 
\begin{subequations}
\label{eq:lambda}
\begin{align}
&  \frac{\delta_{n_{\max}+1}}{\delta_{n_{\max}}} \approx \frac{2n_{\max}+5}{2n_{\max}+3} \frac{\omega_{n_{\max}+1}}{\omega_{n_{\max}}} \frac{1-\frac{2n_{\max}+1}{2n_{\max}+5} \frac{\omega_{n_{\max}+2}}{\omega_{n_{\max}+1}}}{1-\frac{2n_{\max}-1}{2n_{\max}+3} \frac{\omega_{n_{\max}+1}}{\omega_{n_{\max}}}} \\ &  \frac{f_{l,n_{\max}+1}}{f_{l,n_{\max}}} \approx \frac{2(n_{\max}+l) +3}{2(n_{\max}+l)+1} \frac{\omega_{n_{\max}+l}}{\omega_{n_{\max}+l-1}} \times  \\ & \ \ \ \ \ \ \ \ \ \times  \frac{1-\frac{2(n_{\max}+l)-1}{2(n_{\max}+l)+3} \frac{\omega_{n_{\max}+l+1}}{\omega_{n_{\max}+l}}}{1-\frac{2(n_{\max}+l)-3}{2(n_{\max}+l)+1} \frac{\omega_{n_{\max}+l}}{\omega_{n_{\max}+l-1}}} \nonumber \\ & \forall \ l \geq 1 \nonumber
\end{align}
\end{subequations}
Furthermore, as explained in Appendix \ref{sec:app}, both the $l$ and $n$ expansions are controlled by the parameter $x=k\Tau$, with $\Tau$ the neutrino horizon (average comoving distance traveled by neutrino particles through cosmic history, see Eq.(\ref{eq:tau}) and comments below Eq.(\ref{eq:abc}), along with the plot in Fig. \ref{fig:tau}). Specifically, if one wishes to follow the neutrino transfer function up to a time $x$, we found that 
\begin{subequations}
\label{eq:trunc}
\begin{align}
&	l_{\max} \approx \frac{x}{2} \\ & n_{\max} \approx \frac{x^{1.6}}{5}
\end{align}
\end{subequations}	
are approximately sufficient for convergence, up to $x=30$. These are plotted in Fig. \ref{fig:lmaxnmax}. This concludes our discussion on the truncation scheme.

Now we move on to setting the initial conditions for the GBH, starting at early times when all the individual neutrinos still move at the speed of light. Notice that in the relativistic regime $\frac{q}{\epsilon} \to 1$ and all higher velocity weight integrals approach one another: We recover the usual hierarchy of equations for radiation. One can then set the (say adiabatic) initial conditions for the $n=0$ fluid properties as usual \cite{astro-ph/9506072} 
\begin{subequations}
\label{eq:ics}
\begin{align}
	& \delta = -2\psi \\ & \theta = \frac{1}{2} (k^2 \tau)\psi \\ & \sigma = \frac{1}{15}(k\tau)^2 \psi \\ & f_{l,0} = 0 \ \forall l>2.
\end{align}
\end{subequations} 
Along with $f_{l,n} = f_{l,0} \ \forall n>0$ to set the initial condition for the remaining higher velocity weight fluid properties.

In the nonrelativistic regime, these variables get suppressed by powers of $(q/\epsilon)^{2} = v^2 \sim (T_{0}/ma)^2$, and similarly $n=0$ should suffice for most applications. During the transition, however, higher order contributions become important, and must be included for an accurate computation of the ncdm transfer functions.

In the standard approach for including massive neutrinos in the computation of cosmological observables in linear theory, Eq.(\ref{eq:eqm}) are solved for a given number $N_{q}$ of momentum bins, and up to some $l_{\max}$, i.e. a system of dimensionality $N_{q} \times (l_{\max}+1)$. The solution is then used to perform the $q$ integrals in Eq.(\ref{eq:emt}) for the neutrino fluid properties, which in turn are coupled to the perturbations to the other components in the universe via the Einstein equations. This is known to be a computationally slow procedure \cite{astro-ph/9506072, astro-ph/0203507}. Originally, these $q$ integrals were evaluated with a fixed grid of many equally spaced samples. A significant acceleration in the integration procedure was achieved once a kernel-weighted sampling scheme was introduced, allowing for a much smaller number of momentum bins $N_{q}$ (now employed in both CLASS and CAMB \cite{1104.2935, 1201.3654}). 

The GBH is simpler than the standard approach in that it removes the intermediate step of performing the $q$ integrals, i.e. the neutrino fluid properties are directly coupled to the perturbations to the other species in the universe. It is then plausible that the GBH may be faster than the standard approach (a performance comparison between the GBH and the standard approach is left to future work, see Sec. \ref{sec:dc}). However, it could be the case that the dimensionality of the GBH, i.e. $(l_{\max}+1) \times (n_{\max}+1)$ is significantly bigger than $N_{q} \times (l_{\max}+1)$ (and it is actually what happens on the small scales, given the rapid increase of $n_{\max}$ with $x$, as seen in Fig. \ref{fig:lmaxnmax}), for the same achieved accuracy. Notwithstanding, in some cases a very large number of momentum bins is actually necessary, e.g. to accurately obtain the effective sound speed \cite{1712.03944}. The GBH is not plagued with the same issue since all momentum dependence is integrated out of the dynamical equations.

We now have everything we need, i.e. a closed system of dynamical equations plus suitable initial conditions, to consider the numerical implementation of the GBH. This will allow us to compare it with the Boltzmann solvers CLASS and CAMB.

\section{Numerical Implementation}
\label{sec:nr}

As an example of the numerical implementation of the GBH, we will first consider an individual neutrino component, with varying mass: $m=0.02$eV, $m=0.1$eV and $m=0.5$eV. Also, we implement the GBH at intermediate scales compared to the neutrino horizon today, i.e. $k=x/\Tau$ with $x=15$ and $x=30$. On larger scales ($x \lesssim 1$), neutrino velocities are unimportant and a simple viscous fluid approximation suffices i.e. $l_{\max}=2,n_{\max}=0$ should be enough, while on smaller scales (say $x > 30$) accurately obtaining the neutrino transfer functions is not so important because of free-streaming: Neutrino perturbations get washed out and have a negligible impact on matter perturbations. Nonlinear effects also start to kick in. Furthermore, because of the rapid increase of $n_{\max}$ with $x$ found in Fig. \ref{fig:lmaxnmax}, integration time also rapidly increases with $x$.

The neutrino density contrast transfer function, as a function of the scale factor, obtained from the GBH is compared to the output from CLASS. Boltzmann solvers do not produce accurate neutrino transfer functions at their default precision settings, as the codes are tailored to accurately produce the matter power spectrum, and relic neutrinos only have a subleading impact on this observable. To obtain accurate results, we follow the improved settings found in Appendix B of \cite{1712.03944}: Turn off the CLASS ncdm fluid approximation, use a quadrature strategy to perform the $q$-integrals, with $N_{q}=30$ momentum bins, and set $l_{\max}=30$, i.e. a system of 930 equations. We then expect to get subpercent level accuracy for all redshifts and scales of interest. Note that CAMB does not provide the option of outputting neutrino transfer functions as a function of redshift, for a given fixed scale. This is why a direct comparison of the GBH with CAMB is not included at this stage. However, in high precision settings the neutrino transfer functions from CLASS and CAMB are known to agree to a percent level \cite{1712.03944}.

We develop a toy code, where the sources $\phi(a)$ and $\psi(a)$, as given by CLASS, are used to evolve the GBH. In that way, we do not need to solve the Einstein equations, i.e. we do not need to consider the dynamics of perturbations to the other components in the universe, and can evolve the neutrino perturbations alone. The full problem of implementing the GBH in a Boltzmann code, followed by a comparison of performance with the standard method, is beyond the scope of this paper, and is left to future work. This is discussed in Sec. \ref{sec:dc}. Our goal here is to demonstrate that the GBH indeed can be used to produce accurate neutrino transfer functions, i.e. to provide a proof-of-principle for the GBH.

The initial conditions are set, according to Eq.(\ref{eq:ics}), at some arbitrary early time, when all modes of interest are in superhorizon scales, and neutrinos are relativistic. Our choices of $l_{\max}$ and $n_{\max}$ are guided by Eq.(\ref{eq:trunc}). For $x=15$, we set $l_{\max} = 8$ and $n_{\max} = 16$, while for $x=30$, we choose $l_{\max}=15$ and $n_{\max}=49$. 

\begin{figure*}
		\centering
		\includegraphics[width=1\textwidth]{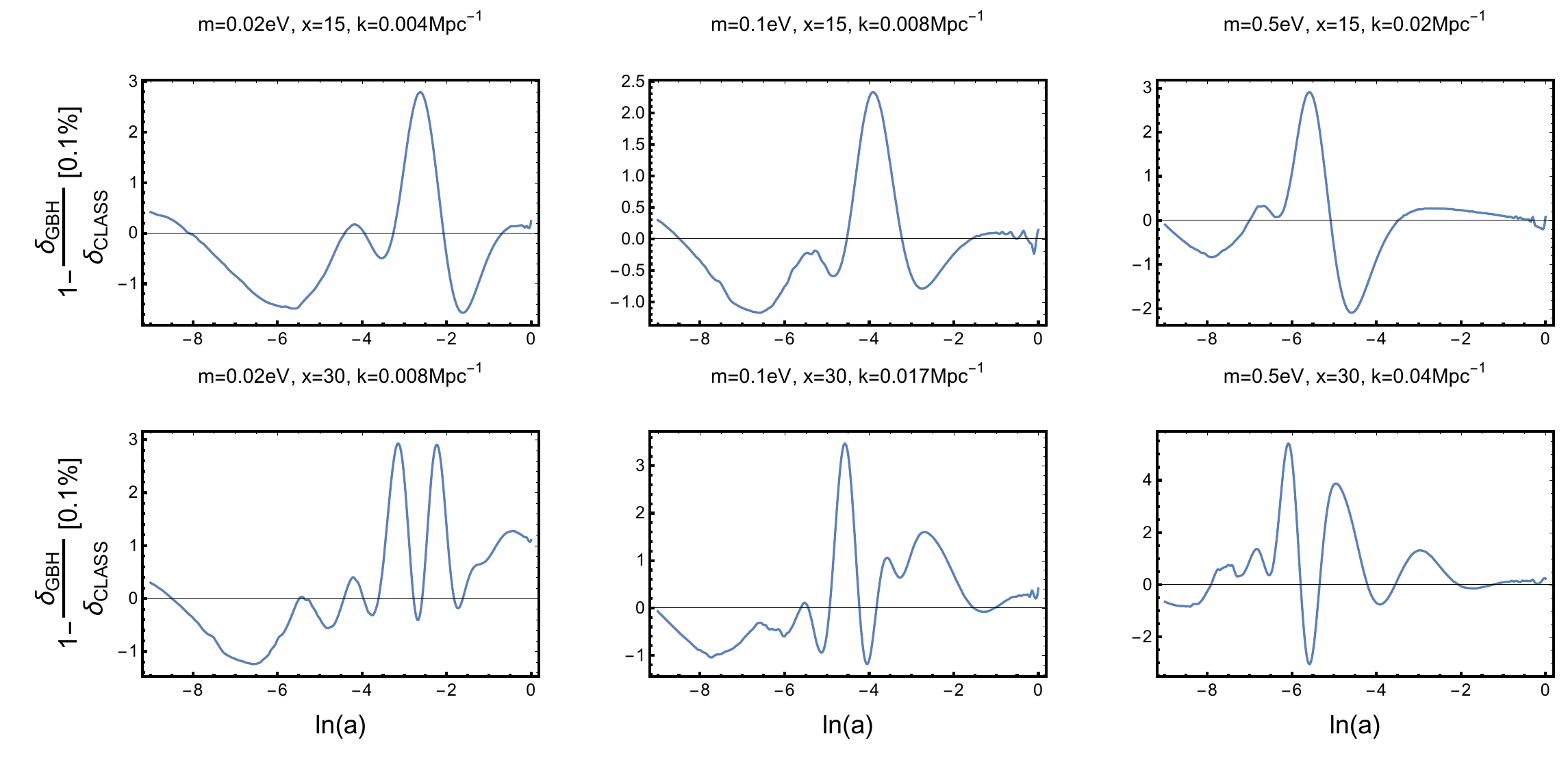}
		\caption{Relative difference in the density contrast neutrino transfer function from the GBH and CLASS (in high precision settings) for neutrino masses of $m=0.02$eV, $m=0.1$eV, and $m=0.5$eV, and at intermediate scales defined by $k=x/\Tau$, with $x=15$ (top) and $x=30$ (bottom). The agreement is in the $\lesssim 0.5\%$ level for all redshift. We conclude that the GBH is accurately producing neutrino transfer functions.}
\label{fig:GFEfull}
\end{figure*}

The relative difference in the transfer functions generated from CLASS and the GBH are shown in Fig. \ref{fig:GFEfull}: There is $\lesssim 0.5\%$ agreement for all redshift, so the GBH is accurately producing neutrino transfer functions. 

The case $x>30$ requires a much larger system of equations to obtain subpercent level agreement with CLASS, with the $n$ expansion converging rather slowly on the small scales. There are also some technical issues with the truncation scheme for the GBH in this regime (see the discussion at the end of Appendix \ref{sec:app}), so going beyond $x=30$ requires a more efficient numerical implementation. 

\begin{figure*}
		\centering
		\includegraphics[width=0.8\textwidth]{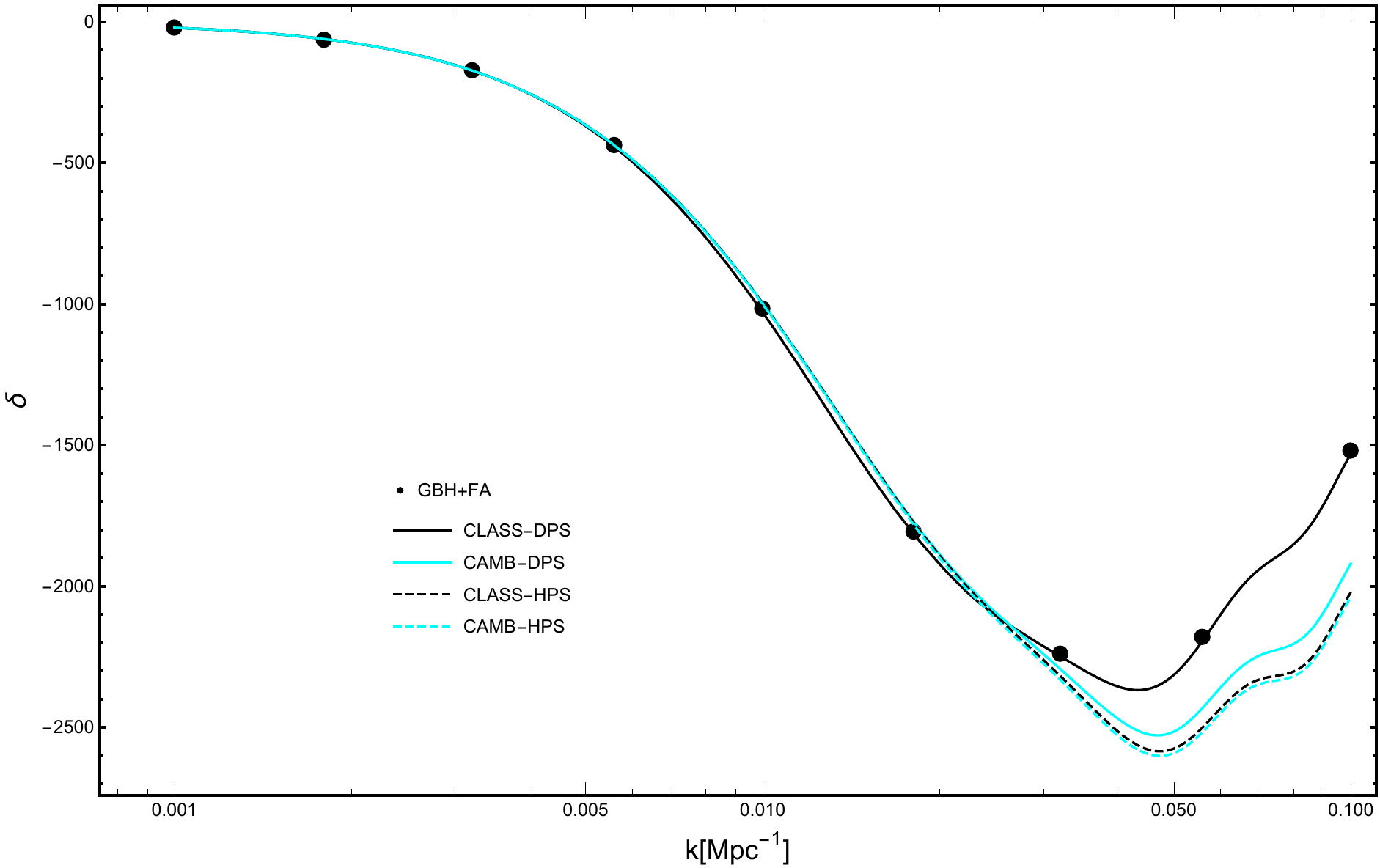}
		\caption{Density contrast neutrino transfer functions, as a function of scale and at redshift $z=0$, from the GBH while switching to a fluid approximation at $x_{\max}=15$ (GBH + FA), CLASS in both default (CLASS-DPS) and high (CLASS-HPS) precision settings, and CAMB in both default (CAMB-DPS) and high (CAMB-HPS) precision settings as well. Here we set the neutrino mass to $m=0.1$eV. We verified that choosing a value of $x_{\max}=30$ for the turning point between the GBH and the FA produced very similar results for the black dots. Also, the neutrino transfer function from CAMB is originally in the synchronous gauge, so we had to perform a gauge transformation to the Newtonian gauge to produce the curves in cyan. }
\label{fig:gfevsdclass}
\end{figure*}

Notwithstanding, and as pointed out before, on the small scales neutrinos free-stream, so it no longer becomes important that neutrino transfer functions are obtained very accurately (nonlinearities also start to kick in). In fact, we find that when switching to a viscous fluid approximation once a given mode is sufficiently inside the horizon (GBH+FA), we produce the density contrast neutrino transfer function as accurately as CLASS in its default precision settings (CLASS-DPS, where a similar switch to a viscous fluid approximation is also employed), over all scales of interest, and at redshift $z=0$. We also find that CAMB in its default precision settings (CAMB-DPS) is as accurate as CLASS-DPS and the GBH+FA at intermediate and large scales, but more accurate on the small scales, where both the GBH+FA and CLASS-DPS are in the FA regime, and hence not producing neutrino transfer functions very accurately. Finally, CAMB in its high-precision settings (CAMB-HPS) \footnote{CAMB-HPS is defined by the following choice of precision settings: massive\_nu\_approx=0, accurate\_massive\_neutrino\_transfers = T, accuracy\_boost=3 and l\_accuracy\_boost=3.} is found to closely agree with CLASS in its high-precision settings (CLASS-HPS), according to expectation. These results are illustrated in Fig. \ref{fig:gfevsdclass}.

\begin{figure*}
	\centering
	\includegraphics[width=0.8\textwidth]{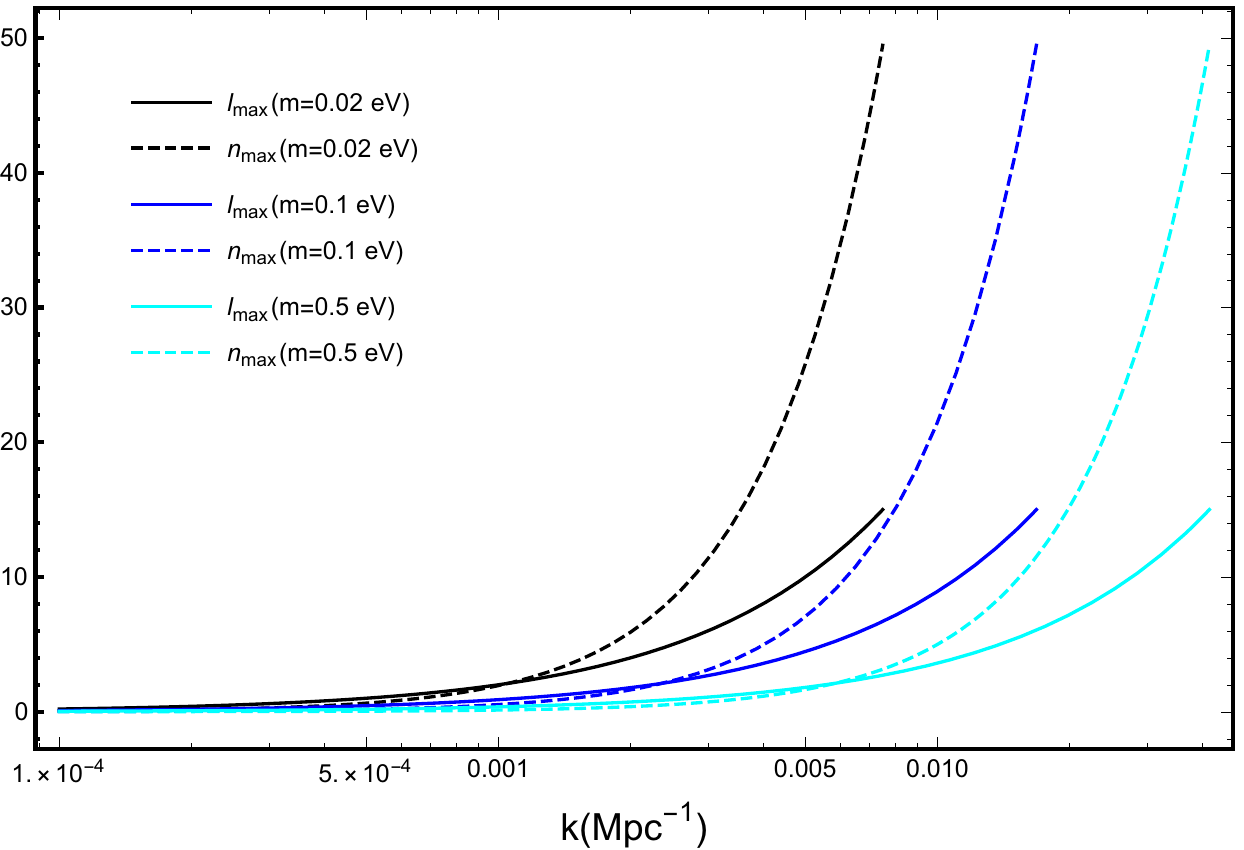}
	\caption{Choices of $l_{\max}$ and $n_{\max}$, as given by Eq.(\ref{eq:trunc}) to approximately ensure convergence of the GBH, as a function of scale for varying neutrino mass, and up to $x=30$. For a given fixed scale, the dimension of the GBH increases for smaller neutrino mass. In fact, since $\Tau \sim 1/\sqrt{m}$, we can use Eq.(\ref{eq:trunc}) to conclude that $l_{\max} \sim 1/\sqrt{m}$, while $n_{\max} \sim 1/m^{0.8}$. }
	\label{fig:lmaxnmaxvsk}
\end{figure*}

Finally, as derived in Appendix \ref{sec:app}, $\Tau \sim \sqrt{T_{0}/m}$ in the nonrelativistic regime. It implies that, and for a given fixed scale $k$, one needs higher $l_{\max}$ and $n_{\max}$ for smaller neutrino masses (assuming that $m$ is big enough for the transition to the nonrelativistic regime to happen before today). In Fig. \ref{fig:lmaxnmaxvsk}, we plot both $l_{\max}$ and $n_{\max}$, as given by Eq.(\ref{eq:trunc}), as a function of scale for varying mass.

\section{Conclusion}
\label{sec:dc}

We introduced the so-called generalized Boltzmann hierarchy (GBH) for noncold dark matter cosmological perturbations in a flat universe, an alternative to the usual Boltzmann hierarchy for accurately producing neutrino transfer functions in the linear regime. It was determined that the GBH agrees with Boltzmann solvers in high precision settings, to a $\lesssim 0.5\%$ level accuracy, in both large and intermediate scales compared to the neutrino free-streaming scale. 

On the small scales one needs to choose a very high $n_{\max}$ in order to produce accurate neutrino transfer functions, and the numerical integration of the GBH becomes computationally expensive. However, one should keep in mind that on small scales free-streaming effects, and nonlinearities, start to kick in, and hence the less important it becomes that the neutrino transfer functions are produced very accurately, which enables a switch to a fluid approximation once a given mode becomes smaller than the free-streaming scale (GBH+FA).

For a given scale and accuracy goal, the GBH approach involves solving at least roughly the same number of equations as a standard Boltzmann solver would, but completely avoids the inconvenience, and computational challenges, associated with solving the hierarchy to later integrate over momentum space, a procedure which is known to be computationally slow, since the momentum integrals are performed to compute neutrino fluid properties, which in turn are coupled to the perturbations to the other components in the universe via the Einstein equations. This feature makes it plausible that the GBH may be faster than the standard approach. 

In this work, we considered the numerical implementation of the GBH in a simplified scenario, where the sources $\phi(a)$ and $\psi(a)$ are obtained directly from CLASS. In this way, we do not need to solve the Einstein equations, and can evolve neutrino perturbations alone, with the goal of providing a proof-of-principle for the GBH. The full problem of implementing the GBH in a Boltzmann solver (allowing for a comparison of performance between the GBH and the standard approach), is left for future work. 

Being a novel approach to the inclusion of massive neutrinos in the computation of cosmological perturbations in linear theory, the GBH is interesting in its own right. It can provide some insight, and serve as cross check for the standard approach. Moreover, while both CLASS and CAMB are now greatly optimized for speed and performance, it is always worthwhile to explore new numerical approaches. \footnote{This is especially true for the implementation of massive neutrinos, one of the most time-consuming tasks of Boltzmann codes.} This is because cosmological analyses often employ Markov chain Monte Carlo (MCMC) methods to map the likelihood in a multidimensional parameter space, requiring Boltzmann codes to be run multiple times, and thus demanding significant computational resources.

In future work, we plan to implement the GBH+FA in the Boltzmann solver CLASS. Based in our findings, we expect to obtain similar accuracy as in the standard approach in default precision settings, and, we surmise, reduced computational time. Furthermore, with a more efficient numerical implementation of the GBH, we will be able to assess its applicability to produce accurate neutrino transfer functions on the small scales, i.e. for $x>30$. This is a current limitation of the GBH, that stops it from producing neutrino transfer functions as accurately as existing Boltzmann solvers in high precision settings, on the small scales (see Fig. \ref{fig:gfevsdclass}). However, we reinforce that because of neutrino free-streaming, the accuracy currently achieved with the GBH+FA is good enough for the vast majority of projects that involve massive neutrinos.

\acknowledgements
I am grateful to Marilena Loverde for helpful correspondence, guidance, and making comments on multiple drafts. I would also like to thank Julien Lesgourgues for the really nice and valuable comments on a draft.

\appendix
\section{Truncation scheme}
\label{sec:app}

We now look for a well-defined truncation scheme for the GBH: For a given neutrino mass $m$ and scale $k$, we must be able to find values of $l_{\max}$ and $n_{\max}$ for which the GBH accurately produces the neutrino transfer functions, and becomes insensitive to a further increase in these parameters. This is, of course, just the statement of convergence. From our experience with the Boltzmann hierarchy, we expect that convergence with respect to $l_{\max}$ is not hard to achieve. In fact, we found that the following truncation, suggested in \cite{astro-ph/9506072}:
\begin{equation}
\label{eq:ltrunc}
	\Psi_{l_{\max}+1} \approx \frac{(2l_{\max}+1)\epsilon}{qk\tau} \Psi_{l_{\max}} - \Psi_{l_{\max}-1}
\end{equation}
is compatible with the structure of the GBH, and produces good results. Substitution of Eq.(\ref{eq:ltrunc}) into Eq.(\ref{eq:fp}) yields
\begin{equation}
\label{eq:ltruncfp}
	f_{l_{\max}+1,n} \approx (l_{\max}+1)\Big(\frac{1}{k\tau} f_{l_{\max},n} - \frac{l_{\max}}{4l_{\max}^2 -1} f_{l_{\max}-1,n+1}\Big)
\end{equation}
for $l_{\max}>1$. Next we move on to the truncation with respect to $n_{\max}$. First notice that
\begin{equation}
\begin{split}
\label{eq:ntrunc}
	 f_{l,n_{\max}+1}-f_{l,n_{\max}} \sim & \int_{0}^{\infty} dq q^2f_{0}(q) \epsilon \Big(\frac{q}{\epsilon}\Big)^{2n_{\max}+l} \times \\ & \times \Big[1-\Big(\frac{q}{\epsilon}\Big)^2\Big]\Psi_{l}
\end{split}
\end{equation}
This integrand contains a term with the form $f(y) = y^k(1-y^2)$, for $y \to q/\epsilon$. This goes to zero in both the relativistic and nonrelativistic regimes, with a peak in between that goes as $1/k$ for $k \gg 1$. If our truncation scheme is based on finding an approximate expression for Eq.(\ref{eq:ntrunc}), it seems reasonable to assume that convergence will be achieved for high enough $n_{\max}$. We proceed in analogy to what is done in \cite{1104.2935}: Find an approximate expression to
\begin{subequations}
\label{eq:lambda}
\begin{align}
& \Lambda_{0} \defeq \frac{\delta_{n_{\max}+1}}{\delta_{n_{\max}}} = \frac{\int_{0}^{\infty} dq q^2f_{0}(q)\epsilon \Big(\frac{q}{\epsilon}\Big)^{2n_{\max}+2} \Psi_{0}}{\int_{0}^{\infty} dq q^2f_{0}(q)\epsilon \Big(\frac{q}{\epsilon}\Big)^{2n_{\max}} \Psi_{0}} \\ & \Lambda_{l} := \frac{f_{l,n_{\max}+1}}{f_{l,n_{\max}}} = \frac{\int_{0}^{\infty} dq q^2f_{0}(q)\epsilon \Big(\frac{q}{\epsilon}\Big)^{2n_{\max}+l+2} \Psi_{l}}{\int_{0}^{\infty} dq q^2f_{0}(q)\epsilon \Big(\frac{q}{\epsilon}\Big)^{2n_{\max}+l} \Psi_{l}} \\ & \forall \ l \geq 1 \nonumber
\end{align}
\end{subequations}
based on an educated guess on the $q/\epsilon$ dependence of the multipoles $\Psi_{l}$. In order to investigate this carefully, let us go back to the Boltzmann hierarchy in Eq.(\ref{eq:eqm}). After setting 
\be
\label{eq:msplit}
\Psi_{l} = -\frac{d\ln f_{0}}{d\ln q} \tilde{\Psi}_{l}
\ee
and introducing a new (q-dependent) time variable,
\be
\label{eq:tau}
\Tau := \int_{i} d\tau \frac{q}{\epsilon} 
\ee
along with $x=k\Tau$, the Boltzmann hierarchy reads
\begin{subequations}
\label{eq:neweqm}
\begin{align}
		& \frac{d\tilde{\Psi}_{0}}{dx} = - \tilde{\Psi}_{1} + \frac{d\phi}{dx} \\ &  \frac{d\tilde{\Psi}_{1}}{dx} = \frac{1}{3} (\tilde{\Psi}_{0}-2\tilde{\Psi}_{2}) + \frac{1}{3} \tilde{\psi} \\ & \frac{d\tilde{\Psi}_{l}}{dx} = \frac{1}{2l+1}[l\tilde{\Psi}_{l-1} - (l+1)\tilde{\Psi}_{l+1}] 
\end{align}
\end{subequations}

This is the same set of equations one would find for radiation, but in terms of the time parameter $\Tau$, and a (q-dependent) effective gravitational potential $\tilde{\psi} = (\epsilon/q)^2 \psi$. This has two important consequences: First, all dependence on scales is actually encoded in $x=k\Tau$, i.e. horizon crossing is effectively defined by the condition that $k \Tau \sim 1$. Second, the mass dependence is encoded in $x$, but also in the effective gravitational potential, and in the nonrelativistic limit it dominates the right-hand side of the evolution equation for $\tilde{\Psi}_{1}$: This is just the well-known decoupling of $l<2$ from higher multipoles in the nonrelativistic regime. It is then true that
\be
\tilde{\Psi}_{1} \sim \int dx \ \tilde{\psi} \sim \int d\tau \frac{\epsilon}{q} \psi
\ee
and hence $\tilde{\Psi}_{1} \propto \epsilon/q$ to leading order, where we think of expanding $\tilde{\Psi}_{l}$ in a power series on $q/\epsilon$ around the nonrelativistic regime. Substitution of this into Eq.(\ref{eq:neweqm}) now implies that $\tilde{\Psi}_{0} \propto 1 $ and $\tilde{\Psi}_{l} \propto (q/\epsilon)^{l-2}$ for $l \geq 1$, to leading order. This, combined with Eq.(\ref{eq:msplit}), are used on Eq.(\ref{eq:lambda})
\begin{subequations}
\label{eq:lambdatrunc}
\begin{align}
& \Lambda_{0} \approx \frac{\int_{0}^{\infty} dq q^2f_{0}(q) \epsilon \frac{d\ln f_{0}}{d\ln q}  \Big(\frac{q}{\epsilon}\Big)^{2(n_{\max}+1)}}{\int_{0}^{\infty} dq q^2f_{0}(q) \epsilon \frac{d\ln f_{0}}{d\ln q} \Big(\frac{q}{\epsilon}\Big)^{2n_{\max}}} \\ & \Lambda_{l} \approx  \frac{\int_{0}^{\infty} dq q^2f_{0}(q) \epsilon \frac{d\ln f_{0}}{d\ln q}  \Big(\frac{q}{\epsilon}\Big)^{2(n_{\max}+l)}}{\int_{0}^{\infty} dq q^2f_{0}(q) \epsilon \frac{d\ln f_{0}}{d\ln q} \Big(\frac{q}{\epsilon}\Big)^{2(n_{\max}+l-1)}} \ \ l \geq 1.
\end{align}
\end{subequations}
After integration by parts, this can be written solely in terms of the background pressures $P_{n}$ (or equation of state parameters $\omega_{n}$) as follows:
\begin{subequations}
\label{eq:lambdatrunc2}
\begin{align}
& \Lambda_{0} \approx \frac{2n_{\max}+5}{2n_{\max}+3} \frac{\omega_{n_{\max}+1}}{\omega_{n_{\max}}} \frac{1-\frac{2n_{\max}+1}{2n_{\max}+5} \frac{\omega_{n_{\max}+2}}{\omega_{n_{\max}+1}}}{1-\frac{2n_{\max}-1}{2n_{\max}+3} \frac{\omega_{n_{\max}+1}}{\omega_{n_{\max}}}} \\ & \Lambda_{l} \approx \frac{2(n_{\max}+l) +3}{2(n_{\max}+l)+1} \frac{\omega_{n_{\max}+l}}{\omega_{n_{\max}+l-1}} \times  \\ & \ \ \ \ \ \ \ \ \ \times  \frac{1-\frac{2(n_{\max}+l)-1}{2(n_{\max}+l)+3} \frac{\omega_{n_{\max}+l+1}}{\omega_{n_{\max}+l}}}{1-\frac{2(n_{\max}+l)-3}{2(n_{\max}+l)+1} \frac{\omega_{n_{\max}+l}}{\omega_{n_{\max}+l-1}}} \nonumber
\end{align}
\end{subequations}

There is only one final piece of information that needs to be specified in order to complete the truncation scheme: How to choose the values of $l_{\max}$ and $n_{\max}$. We know that higher multipoles and higher velocity weight fluid properties contribute as small scale effects, acting as viscosity, since on large scales a simple fluid approximation suffices. Based on this, and the observations made following Eq.(\ref{eq:neweqm}), we expect that higher values of $l_{\max}$ and $n_{\max}$ are needed as $x=k\Tau$ increases.

Due to its importance, let us stop for a moment to study the time variable $\Tau$, defined in Eq.(\ref{eq:tau}). During the relativistic regime, it is identical to $\tau$. Let us now see what happens in the nonrelativistic regime, assuming that the time of transition $a_{\textrm{tr}} \sim q/m$, happens during matter domination, as is the case for massive neutrinos. We may then write the following approximation:
\be
\label{eq:tauapprox}
\Tau \approx \tau_{\textrm{tr}} + \int_{\textrm{tr}} \frac{da}{a} \frac{1}{aH} \frac{q}{ma}
\ee
where we split the integral from the initial time to the transition, and from the transition to the final time, use $d\tau = \frac{da}{a} \frac{1}{aH}$, with $H= \frac{\mathcal{H}}{a}$ is the Hubble rate, and approximate $\epsilon \approx ma$ in the nonrelativistic regime, along with $\epsilon \approx q$ up to the transition. Further using $H \sim a^{-3/2}$ during matter domination, one obtains for the integral in the right-hand side of Eq.(\ref{eq:tauapprox})
\be
\label{eq:abc}
 \int_{\textrm{tr}} \frac{da}{a} \frac{1}{aH} \frac{q}{ma} \sim \frac{q}{m} \int_{\textrm{tr}}  \frac{da}{a} a^{-\frac{1}{2}} \sim \frac{q}{m} a_{\textrm{tr}}^{-\frac{1}{2}} \sim \sqrt{\frac{q}{m}},
\ee
where we use the fact that the integral is dominated by its lower limit, and use $a_{\textrm{tr}} \sim q/m$. Further notice that during matter domination $\tau_{\textrm{tr}} \sim a_{\textrm{tr}}^{1/2} \sim \sqrt{\frac{q}{m}}$ as well, so $\Tau \sim \sqrt{\frac{T_{0}}{m}}$ approaches a time-independent constant, as opposed to $\tau$, which grows indefinitely. In other words, $\Tau$ grows like $\tau$ up to the time of transition, effectively freezing as one approaches the nonrelativistic regime. Indeed, since $q/\epsilon = v$ is the neutrino velocity, $\Tau$ is the comoving distance traveled by a neutrino particle through cosmic history. When evaluated at the peak of the Fermi-Dirac distribution (say $q/T_{0} =3$), this is roughly the neutrino horizon, or the free-streaming scale (integrated over e-folds). From this point forward, when used as a time variable, it is implicitly assumed that $T$ is evaluated at $q/T_{0} =3$, and hence corresponds to the neutrino horizon (plotted in Fig. \ref{fig:tau}).

On the large scales, i.e. $x=k\Tau \lesssim 1$, a simple viscous fluid approximation with $l_{\max}=2$ and $n_{\max}=0$ suffices, and we discuss this in detail in Appendix \ref{sec:CLASS}. 

As $x \gg 1$, we expect that higher values of both $l_{\max}$ and $n_{\max}$ are needed. A naive assumption would then be that $l_{\max}, n_{\max} \propto x$. Indeed, our experience with the GBH indicates that it converges for $l_{\max} \approx x/2$, with the difference between $\Tau$ and $\tau$ explaining why one needs a much higher $l_{\max}$ for radiation than for massive neutrinos \cite{astro-ph/9506072}. Unfortunately, the same cannot be said about the $n$ expansion, with $n_{\max}$, and hence the dimensionality of the system, growing very rapidly for modes inside the horizon. This can be explained as follows: The higher velocity weight variables are effectively accounting for the dynamics of neutrinos with higher particle velocities. This means that $\Tau$, as defined by Eq.(\ref{eq:tau}), should not be evaluated at the peak of the Fermi-Dirac distribution $q/T_{0} \approx 3$, but rather at a larger value $q=q_{n_{\max}}$, that we choose to be the peak of the integrand in Eq.(\ref{eq:bfp}), for the corresponding value of $n=n_{\max}$. This is to make sure that $\Tau(q_{n_{\max}})$ is the comoving distance traveled by the neutrino particles that are actually being probed by the higher-velocity weight fluid properties. We then expect that
\begin{equation}
\label{eq:nmaxvsk}
	\frac{n_{\max}}{x} \sim \frac{k\Tau(q_{n_{\max}})}{x} = \frac{\Tau(q_{n_{\max}})}{\Tau}  
\end{equation}

We evaluated Eq.(\ref{eq:nmaxvsk}) numerically for many values of $n_{\max}$, to find that $ n_{\max} \sim x^{1.6}$ up to $x=30$. From our experience with the GBH, $n_{\max} \approx x^{1.6}/5$ is approximately sufficient for convergence.

Before moving on to the viscous fluid approximation, we should point out that our truncation scheme in Eq. \ref{eq:lambdatrunc2} requires accurate numerical evaluation of quotients involving higher-velocity weight equations of state, which according to Eq.(\ref{eq:bfp}), become really small numbers in the nonrelativistic regime. The numerical computation of $\Lambda_{l}$ can then be very time consuming, especially for high $n_{\max}$. However, as it is clear from Eq.(\ref{eq:bfp}), $\omega_{n}$ is a function of just a single variable $y= ma/T_{0}$, and hence can be easily tabulated.

\section{Fluid approximation}
\label{sec:CLASS}

On large scales for which $x=k\Tau \lesssim 1$, a simple viscous fluid approximation with $l_{\max}=2$ and $n_{\max}=0$ should suffice. Since in the nonrelativistic regime $\Tau$ basically freezes at $\tau_{\textrm{tr}}$, the value of conformal time evaluated at the transition $a_{\textrm{tr}} \sim \frac{q}{m} \sim \frac{T_{0}}{m}$, this can be rephrased as to say that the mode has to be superhorizon at the transition, i.e. in enters the horizon during the nonrelativistic regime. This is exactly what was found in \cite{1003.0942}, from comparing the exact solution with a simple fluid approximation. 

Furthermore, on the small scales $x \gg 1$, free-streaming and nonlinear effects start to kick in and it no longer becomes important that the neutrino transfer function is produced very accurately. One can then choose a $x_{\max}$ above which a simple fluid approximation can be used once again. Let us then stop for a moment to carefully study the case $l_{\max}=2$ and $n_{\max}=0$.  The fluid equations are given by Eq.(\ref{eq:ncdmfe}). Our truncation scheme (TS), developed in Appendix \ref{sec:app}, provides approximations for the quantities  $\delta P/\delta \rho$, $\Sigma$, $\Theta$ and $kf_{3}$, in terms of the dynamical variables in the system. 

Setting $l_{\max}=2$ and $n_{\max}=0$ in Eq.(\ref{eq:ltruncfp}) (which is known not to be a particularly good approximation for $l_{\max}=2$ \cite{1510.02907}), one obtains 
\be
\label{eq:mtrunc}
kf_{3} \approx \frac{3}{\tau} \sigma - \frac{2}{5} \Theta
\ee

\begin{figure}
		\centering
		\includegraphics[width=0.5\textwidth]{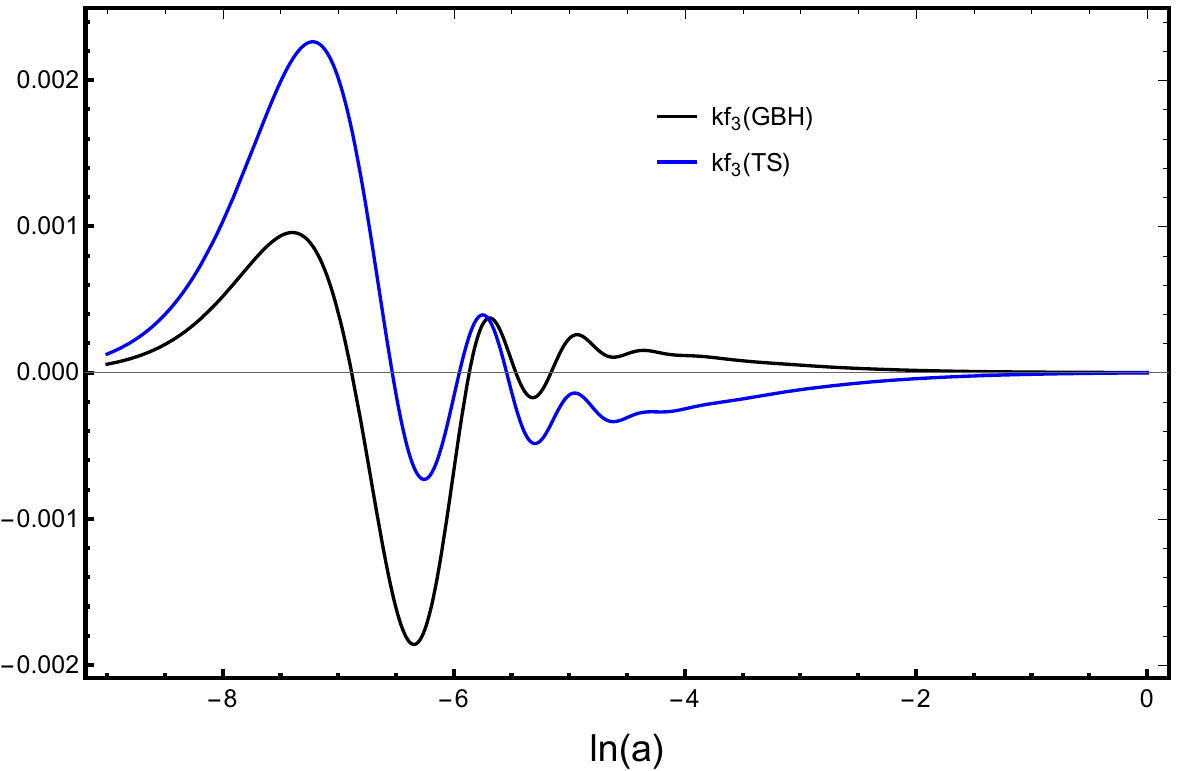}
		\caption{Left- (GBH) and right-hand (TS) sides of Eq.(\ref{eq:mtrunc}) as obtained from the GBH. Here $x=15$, or $k=0.008$Mpc$^{-1}$, $m = 0.1$eV, $l_{\max} = 8$, and $n_{\max} = 16$. The approximation reproduces the right features, but is not accurate.}
\label{fig:f3}
\end{figure}

Both sides of this equation can be accurately determined at intermediate scales, from the GBH. A comparison can be found in Fig. \ref{fig:f3}.

Furthermore, set $n_{\max}=0$ into Eq.(\ref{eq:lambdatrunc2}) to arrive at
\begin{subequations}
\label{eq:hmt}
\begin{align}
	\frac{1}{3} \Lambda_{0} = \frac{\delta P}{\delta \rho} & \approx \frac{5}{3} \frac{\omega}{1+\omega} \big(1-\frac{1}{5}\frac{\omega_{2}}{\omega_{1}}\big) = c_{g}^{2} \\ \Lambda_{1} = \frac{\Theta}{\theta} &\approx 3c_{g}^{2} \iff c_{\textrm{vis}}^2 = \frac{3}{4}(1+\omega)c_{g}^2 \\ \Lambda_{2} = \frac{\Sigma}{\sigma} & \approx \frac{7}{5} \frac{\omega_{2}}{\omega_{1}} \frac{1-\frac{1}{7} \frac{\omega_{3}}{\omega_{2}}}{1-\frac{1}{5} \frac{\omega_{2}}{\omega_{1}}}
\end{align}
\end{subequations}
where a fluid viscosity speed $c_{\textrm{vis}}$, was introduced as a different parametrization to $\Lambda_{1}$, using notation from \cite{astro-ph/9801234}
\be
\label{eq:vis}
c_{\textrm{vis}}^2 = \frac{1}{4}(1+\omega)\frac{\Theta}{\theta}
\ee

In the CLASS ncdm fluid approximation, the truncation in the multipole is done in the exact same way as in Eq.(\ref{eq:mtrunc}), while for the higher velocity weight quantities, the authors of \cite{1104.2935} apply a bit of trial and error to arrive at the following \textit{ad hoc} approximations
\begin{subequations}
\label{eq:hmt2}
\begin{align}
	\frac{1}{3} \Lambda_{0} = \frac{\delta P}{\delta \rho} & \approx \frac{5}{3} \frac{\omega}{1+\omega} \big(1-\frac{1}{5}\frac{\omega_{2}}{\omega_{1}}\big) = c_{g}^{2} \\ \Lambda_{1} =\frac{\Theta}{\theta} &\approx 12 \frac{\omega}{1+\omega} c_{g}^{2} \iff c_{\textrm{vis}}^2 = 3\omega c_{g}^{2} \\ \Lambda_{2} =\frac{\Sigma}{\sigma} & \approx  \frac{\omega_{2}}{\omega_{1}}
\end{align}
\end{subequations}

The Eqs.(\ref{eq:hmt}) and (\ref{eq:hmt2}) differ slightly on the expressions for  $\Theta/\theta$ and $\Sigma/\sigma$. 

\begin{figure}
		\centering
		\includegraphics[width=0.5\textwidth]{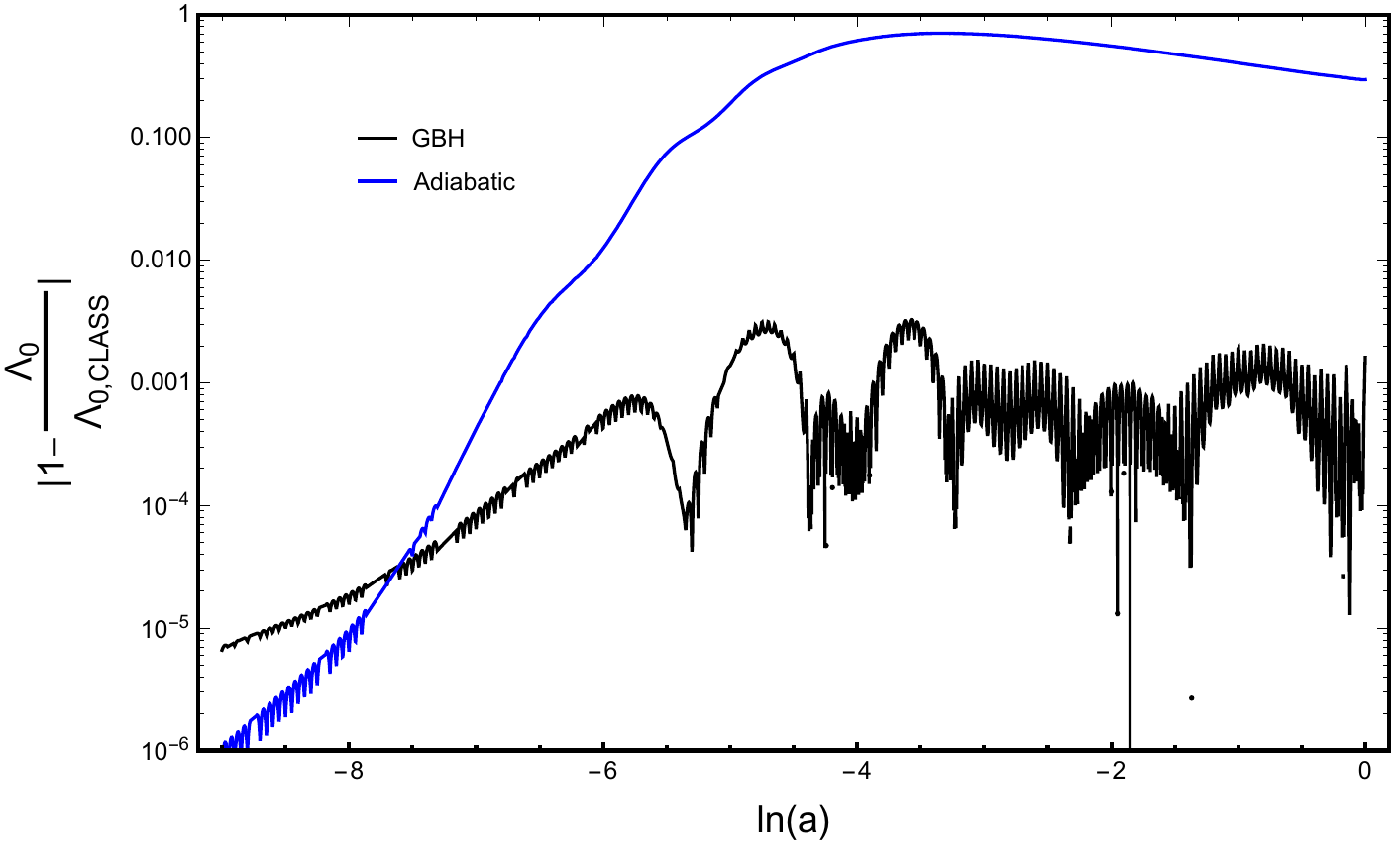}
		\caption{Relative difference between sound speeds squared, coming from the GBH and the assumption of adiabaticity, when compared to the exact solution from CLASS. Here $x=15$, or $k=0.008$Mpc$^{-1}$, $m = 0.1$eV, $l_{\max} = 8$, and $n_{\max} = 16$.}
\label{fig:cs2}
\end{figure}

In Fig. \ref{fig:cs2} we compare the adiabatic sound speed squared to the exact solutions coming from both CLASS and the GBH: The GBH and CLASS agree to a subpercent level, with both differing from the assumption of adiabaticity when approaching the nonrelativistic regime.

\begin{figure}
		\centering
		\includegraphics[width=0.5\textwidth]{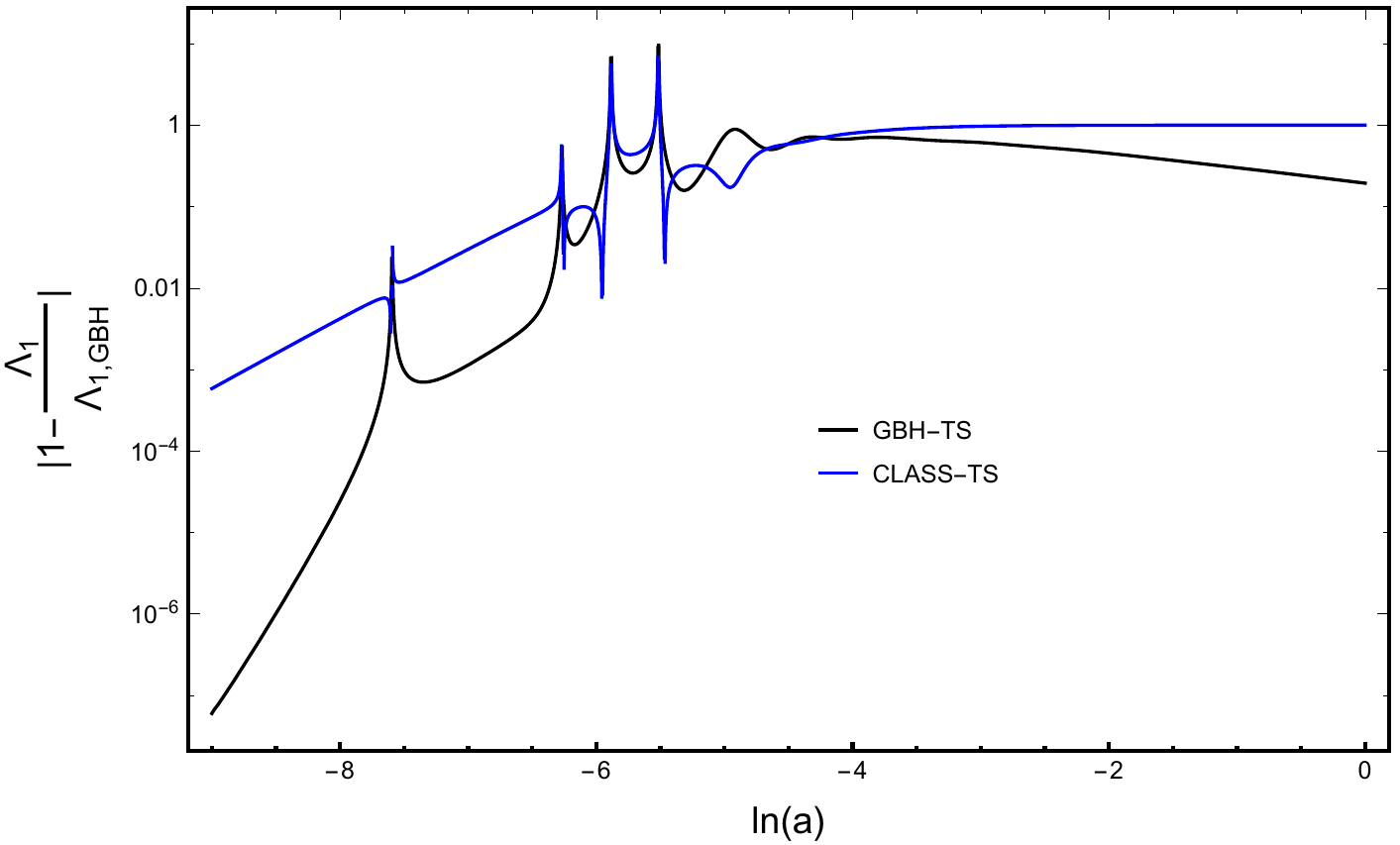}
		\caption{Relative difference between $\Lambda_{1}$'s, coming from CLASS and GBH truncation schemes, when compared to the exact solution from the GBH. Here $x=15$, or $k=0.008$Mpc$^{-1}$, $m = 0.1$eV, $l_{\max} = 8$, and $n_{\max} = 16$.}
\label{fig:lambda1}
\end{figure}

In Fig. \ref{fig:lambda1}, we compare the $\Lambda_{1}$'s from CLASS and GBH truncation schemes for the FA with the exact solution from the GBH: There is an overall $10\%$ level error in both cases, but the GBH truncation scheme is an order of magnitude better in the nonrelativistic regime. 

\begin{figure}
		\centering
		\includegraphics[width=0.5\textwidth]{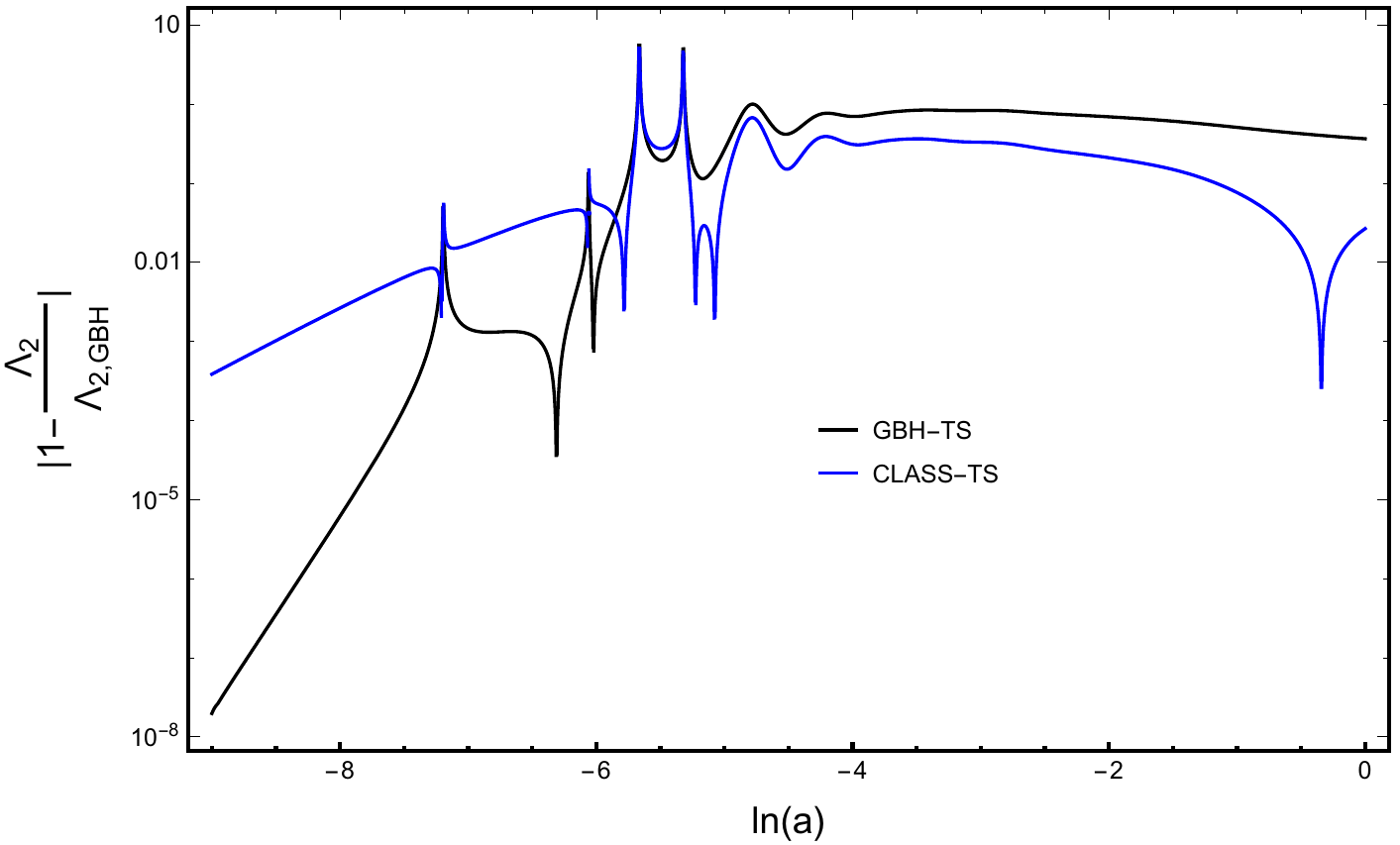}
		\caption{Relative difference between $\Lambda_{2}$'s, coming from CLASS and GBH truncation schemes, when compared to the exact solution from the GBH. Here $x=15$, or $k=0.008$Mpc$^{-1}$, $m = 0.1$eV, $l_{\max} = 8$, and $n_{\max} = 16$.}
\label{fig:lambda2}
\end{figure}

In Fig. \ref{fig:lambda2}, we compare $\Lambda_{2}$'s from CLASS and GBH truncation schemes for the FA with the exact solution from the GBH: there is a $10\%$ level error in both cases, with the CLASS truncation scheme being an order of magnitude better in the nonrelativistic regime. However, we found that the accuracy of the FA is insensitive to the specific choice of $\Lambda_{2}$. 

\begin{figure}[H]
		\centering
		\includegraphics[width=0.5\textwidth]{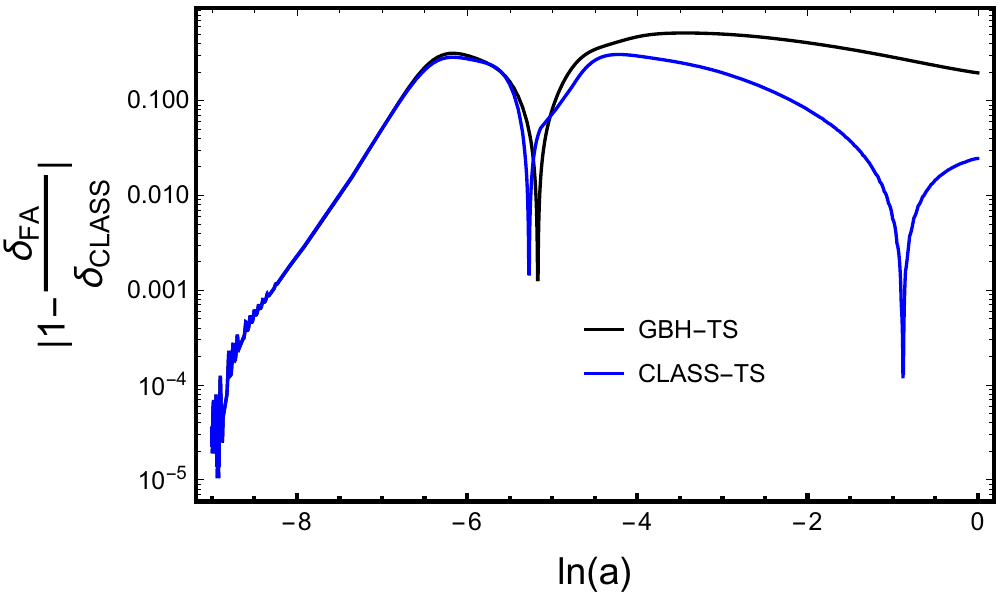}
		\caption{Relative difference between density contrast neutrino transfer functions coming from the fluid approximation with GBH and CLASS truncation schemes, when compared to the exact solution from CLASS. Here $x=15$, or $k=0.008$Mpc$^{-1}$, and $m = 0.1$eV.}
\label{fig:CFA}
\end{figure}

Finally, in Fig. \ref{fig:CFA} we compare fluid approximations, with GBH and CLASS truncation schemes, with the exact solution from CLASS: there is also an overall $10\%$ level error in the neutrino density contrast transfer function, but the CLASS fluid approximation works better at late times. Indeed, this is possible because the CLASS truncation scheme is tuned to produce the best outcome, even though all the individual approximations are in a similar level of accuracy as in the truncation scheme for the GBH, which was motivated from first principles. Because of this, the CLASS fluid approximation works better overall, and should be the one used in the regime $x>x_{\max}$, as discussed.

\bibliography{GFE.bib}

\begin{thebibliography}{30}%
\makeatletter
\providecommand \@ifxundefined [1]{%
 \@ifx{#1\undefined}
}%
\providecommand \@ifnum [1]{%
 \ifnum #1\expandafter \@firstoftwo
 \else \expandafter \@secondoftwo
 \fi
}%
\providecommand \@ifx [1]{%
 \ifx #1\expandafter \@firstoftwo
 \else \expandafter \@secondoftwo
 \fi
}%
\providecommand \natexlab [1]{#1}%
\providecommand \enquote  [1]{``#1''}%
\providecommand \bibnamefont  [1]{#1}%
\providecommand \bibfnamefont [1]{#1}%
\providecommand \citenamefont [1]{#1}%
\providecommand \href@noop [0]{\@secondoftwo}%
\providecommand \href [0]{\begingroup \@sanitize@url \@href}%
\providecommand \@href[1]{\@@startlink{#1}\@@href}%
\providecommand \@@href[1]{\endgroup#1\@@endlink}%
\providecommand \@sanitize@url [0]{\catcode `\\12\catcode `\$12\catcode
  `\&12\catcode `\#12\catcode `\^12\catcode `\_12\catcode `\%12\relax}%
\providecommand \@@startlink[1]{}%
\providecommand \@@endlink[0]{}%
\providecommand \url  [0]{\begingroup\@sanitize@url \@url }%
\providecommand \@url [1]{\endgroup\@href {#1}{\urlprefix }}%
\providecommand \urlprefix  [0]{URL }%
\providecommand \Eprint [0]{\href }%
\providecommand \doibase [0]{http://dx.doi.org/}%
\providecommand \selectlanguage [0]{\@gobble}%
\providecommand \bibinfo  [0]{\@secondoftwo}%
\providecommand \bibfield  [0]{\@secondoftwo}%
\providecommand \translation [1]{[#1]}%
\providecommand \BibitemOpen [0]{}%
\providecommand \bibitemStop [0]{}%
\providecommand \bibitemNoStop [0]{.\EOS\space}%
\providecommand \EOS [0]{\spacefactor3000\relax}%
\providecommand \BibitemShut  [1]{\csname bibitem#1\endcsname}%
\let\auto@bib@innerbib\@empty
%</preamble>
\bibitem [{\citenamefont {de~Salas}\ \emph {et~al.}(2018)\citenamefont
  {de~Salas}, \citenamefont {Forero}, \citenamefont {Ternes}, \citenamefont
  {Tortola},\ and\ \citenamefont {Valle}}]{1708.01186}%
  \BibitemOpen
  \bibfield  {author} {\bibinfo {author} {\bibfnamefont {P.~F.}\ \bibnamefont
  {de~Salas}}, \bibinfo {author} {\bibfnamefont {D.~V.}\ \bibnamefont
  {Forero}}, \bibinfo {author} {\bibfnamefont {C.~A.}\ \bibnamefont {Ternes}},
  \bibinfo {author} {\bibfnamefont {M.}~\bibnamefont {Tortola}}, \ and\
  \bibinfo {author} {\bibfnamefont {J.~W.~F.}\ \bibnamefont {Valle}},\ }\href
  {\doibase 10.1016/j.physletb.2018.06.019} {\bibfield  {journal} {\bibinfo
  {journal} {Phys. Lett. B}\ }\textbf {\bibinfo {volume} {782}},\ \bibinfo
  {pages} {633} (\bibinfo {year} {2018})},\ \Eprint
  {http://arxiv.org/abs/1708.01186} {arXiv:1708.01186 [hep-ph]} \BibitemShut
  {NoStop}%
\bibitem [{\citenamefont {De~Salas}\ \emph {et~al.}(2018)\citenamefont
  {De~Salas}, \citenamefont {Gariazzo}, \citenamefont {Mena}, \citenamefont
  {Ternes},\ and\ \citenamefont {T\'ortola}}]{1806.11051}%
  \BibitemOpen
  \bibfield  {author} {\bibinfo {author} {\bibfnamefont {P.~F.}\ \bibnamefont
  {De~Salas}}, \bibinfo {author} {\bibfnamefont {S.}~\bibnamefont {Gariazzo}},
  \bibinfo {author} {\bibfnamefont {O.}~\bibnamefont {Mena}}, \bibinfo {author}
  {\bibfnamefont {C.~A.}\ \bibnamefont {Ternes}}, \ and\ \bibinfo {author}
  {\bibfnamefont {M.}~\bibnamefont {T\'ortola}},\ }\href {\doibase
  10.3389/fspas.2018.00036} {\bibfield  {journal} {\bibinfo  {journal} {Front.
  Astron. Space Sci.}\ }\textbf {\bibinfo {volume} {5}},\ \bibinfo {pages} {36}
  (\bibinfo {year} {2018})},\ \Eprint {http://arxiv.org/abs/1806.11051}
  {arXiv:1806.11051 [hep-ph]} \BibitemShut {NoStop}%
\bibitem [{\citenamefont {Gerbino}(2018)}]{1803.11545}%
  \BibitemOpen
  \bibfield  {author} {\bibinfo {author} {\bibfnamefont {M.}~\bibnamefont
  {Gerbino}},\ }in\ \href@noop {} {\emph {\bibinfo {booktitle} {{Prospects in
  Neutrino Physics}}}}\ (\bibinfo {year} {2018})\ \Eprint
  {http://arxiv.org/abs/1803.11545} {arXiv:1803.11545 [astro-ph.CO]}
  \BibitemShut {NoStop}%
\bibitem [{\citenamefont {Lesgourgues}\ and\ \citenamefont
  {Pastor}(2006)}]{astro-ph/0603494}%
  \BibitemOpen
  \bibfield  {author} {\bibinfo {author} {\bibfnamefont {J.}~\bibnamefont
  {Lesgourgues}}\ and\ \bibinfo {author} {\bibfnamefont {S.}~\bibnamefont
  {Pastor}},\ }\href {\doibase 10.1016/j.physrep.2006.04.001} {\bibfield
  {journal} {\bibinfo  {journal} {Phys. Rept.}\ }\textbf {\bibinfo {volume}
  {429}},\ \bibinfo {pages} {307} (\bibinfo {year} {2006})},\ \Eprint
  {http://arxiv.org/abs/astro-ph/0603494} {arXiv:astro-ph/0603494} \BibitemShut
  {NoStop}%
\bibitem [{\citenamefont {Loureiro}\ \emph {et~al.}(2019)\citenamefont
  {Loureiro} \emph {et~al.}}]{1811.02578}%
  \BibitemOpen
  \bibfield  {author} {\bibinfo {author} {\bibfnamefont {A.}~\bibnamefont
  {Loureiro}} \emph {et~al.},\ }\href {\doibase 10.1103/PhysRevLett.123.081301}
  {\bibfield  {journal} {\bibinfo  {journal} {Phys. Rev. Lett.}\ }\textbf
  {\bibinfo {volume} {123}},\ \bibinfo {pages} {081301} (\bibinfo {year}
  {2019})},\ \Eprint {http://arxiv.org/abs/1811.02578} {arXiv:1811.02578
  [astro-ph.CO]} \BibitemShut {NoStop}%
\bibitem [{\citenamefont {Aghanim}\ \emph {et~al.}(2020)\citenamefont {Aghanim}
  \emph {et~al.}}]{1807.06209}%
  \BibitemOpen
  \bibfield  {author} {\bibinfo {author} {\bibfnamefont {N.}~\bibnamefont
  {Aghanim}} \emph {et~al.} (\bibinfo {collaboration} {Planck}),\ }\href
  {\doibase 10.1051/0004-6361/201833910} {\bibfield  {journal} {\bibinfo
  {journal} {Astron. Astrophys.}\ }\textbf {\bibinfo {volume} {641}},\ \bibinfo
  {pages} {A6} (\bibinfo {year} {2020})},\ \Eprint
  {http://arxiv.org/abs/1807.06209} {arXiv:1807.06209 [astro-ph.CO]}
  \BibitemShut {NoStop}%
\bibitem [{\citenamefont {Jha}\ \emph {et~al.}(2019)\citenamefont {Jha} \emph
  {et~al.}}]{1907.08945}%
  \BibitemOpen
  \bibfield  {author} {\bibinfo {author} {\bibfnamefont {S.~W.}\ \bibnamefont
  {Jha}} \emph {et~al.},\ }\href@noop {} {\  (\bibinfo {year} {2019})},\
  \Eprint {http://arxiv.org/abs/1907.08945} {arXiv:1907.08945 [astro-ph.IM]}
  \BibitemShut {NoStop}%
\bibitem [{\citenamefont {Blanchard}\ \emph {et~al.}(2020)\citenamefont
  {Blanchard} \emph {et~al.}}]{1910.09273}%
  \BibitemOpen
  \bibfield  {author} {\bibinfo {author} {\bibfnamefont {A.}~\bibnamefont
  {Blanchard}} \emph {et~al.} (\bibinfo {collaboration} {Euclid}),\ }\href
  {\doibase 10.1051/0004-6361/202038071} {\bibfield  {journal} {\bibinfo
  {journal} {Astron. Astrophys.}\ }\textbf {\bibinfo {volume} {642}},\ \bibinfo
  {pages} {A191} (\bibinfo {year} {2020})},\ \Eprint
  {http://arxiv.org/abs/1910.09273} {arXiv:1910.09273 [astro-ph.CO]}
  \BibitemShut {NoStop}%
\bibitem [{\citenamefont {Sehgal}\ \emph {et~al.}(2019)\citenamefont {Sehgal}
  \emph {et~al.}}]{1906.10134}%
  \BibitemOpen
  \bibfield  {author} {\bibinfo {author} {\bibfnamefont {N.}~\bibnamefont
  {Sehgal}} \emph {et~al.},\ }\href@noop {} {\  (\bibinfo {year} {2019})},\
  \Eprint {http://arxiv.org/abs/1906.10134} {arXiv:1906.10134 [astro-ph.CO]}
  \BibitemShut {NoStop}%
\bibitem [{\citenamefont {Spergel}\ \emph {et~al.}(2013)\citenamefont {Spergel}
  \emph {et~al.}}]{1305.5422}%
  \BibitemOpen
  \bibfield  {author} {\bibinfo {author} {\bibfnamefont {D.}~\bibnamefont
  {Spergel}} \emph {et~al.},\ }\href@noop {} {\  (\bibinfo {year} {2013})},\
  \Eprint {http://arxiv.org/abs/1305.5422} {arXiv:1305.5422 [astro-ph.IM]}
  \BibitemShut {NoStop}%
\bibitem [{\citenamefont {Dawson}\ \emph {et~al.}(2016)\citenamefont {Dawson}
  \emph {et~al.}}]{1508.04473}%
  \BibitemOpen
  \bibfield  {author} {\bibinfo {author} {\bibfnamefont {K.~S.}\ \bibnamefont
  {Dawson}} \emph {et~al.},\ }\href {\doibase 10.3847/0004-6256/151/2/44}
  {\bibfield  {journal} {\bibinfo  {journal} {Astron. J.}\ }\textbf {\bibinfo
  {volume} {151}},\ \bibinfo {pages} {44} (\bibinfo {year} {2016})},\ \Eprint
  {http://arxiv.org/abs/1508.04473} {arXiv:1508.04473 [astro-ph.CO]}
  \BibitemShut {NoStop}%
\bibitem [{\citenamefont {Godfrey}\ \emph {et~al.}(2012)\citenamefont
  {Godfrey}, \citenamefont {Bignall}, \citenamefont {Tingay}, \citenamefont
  {Harvey-Smith}, \citenamefont {Kramer}, \citenamefont {Burke-Spolaor},
  \citenamefont {Miller-Jones}, \citenamefont {Johnston-Hollitt}, \citenamefont
  {Ekers},\ and\ \citenamefont {Gulyaev}}]{1111.6398}%
  \BibitemOpen
  \bibfield  {author} {\bibinfo {author} {\bibfnamefont {L.~E.~H.}\
  \bibnamefont {Godfrey}}, \bibinfo {author} {\bibfnamefont {H.}~\bibnamefont
  {Bignall}}, \bibinfo {author} {\bibfnamefont {S.}~\bibnamefont {Tingay}},
  \bibinfo {author} {\bibfnamefont {L.}~\bibnamefont {Harvey-Smith}}, \bibinfo
  {author} {\bibfnamefont {M.}~\bibnamefont {Kramer}}, \bibinfo {author}
  {\bibfnamefont {S.}~\bibnamefont {Burke-Spolaor}}, \bibinfo {author}
  {\bibfnamefont {J.~C.~A.}\ \bibnamefont {Miller-Jones}}, \bibinfo {author}
  {\bibfnamefont {M.}~\bibnamefont {Johnston-Hollitt}}, \bibinfo {author}
  {\bibfnamefont {R.}~\bibnamefont {Ekers}}, \ and\ \bibinfo {author}
  {\bibfnamefont {S.}~\bibnamefont {Gulyaev}},\ }\href {\doibase
  10.1071/AS11050} {\bibfield  {journal} {\bibinfo  {journal} {Publ. Astron.
  Soc. Austral.}\ }\textbf {\bibinfo {volume} {29}},\ \bibinfo {pages} {42}
  (\bibinfo {year} {2012})},\ \Eprint {http://arxiv.org/abs/1111.6398}
  {arXiv:1111.6398 [astro-ph.IM]} \BibitemShut {NoStop}%
\bibitem [{\citenamefont {Levi}\ \emph {et~al.}(2019)\citenamefont {Levi} \emph
  {et~al.}}]{1907.10688}%
  \BibitemOpen
  \bibfield  {author} {\bibinfo {author} {\bibfnamefont {M.~E.}\ \bibnamefont
  {Levi}} \emph {et~al.} (\bibinfo {collaboration} {DESI}),\ }\href@noop {} {\
  (\bibinfo {year} {2019})},\ \Eprint {http://arxiv.org/abs/1907.10688}
  {arXiv:1907.10688 [astro-ph.IM]} \BibitemShut {NoStop}%
\bibitem [{\citenamefont {Dvorkin}\ \emph {et~al.}(2019)\citenamefont {Dvorkin}
  \emph {et~al.}}]{1903.03689}%
  \BibitemOpen
  \bibfield  {author} {\bibinfo {author} {\bibfnamefont {C.}~\bibnamefont
  {Dvorkin}} \emph {et~al.},\ }\href@noop {} {\  (\bibinfo {year} {2019})},\
  \Eprint {http://arxiv.org/abs/1903.03689} {arXiv:1903.03689 [astro-ph.CO]}
  \BibitemShut {NoStop}%
\bibitem [{\citenamefont {Brinckmann}\ \emph {et~al.}(2019)\citenamefont
  {Brinckmann}, \citenamefont {Hooper}, \citenamefont {Archidiacono},
  \citenamefont {Lesgourgues},\ and\ \citenamefont {Sprenger}}]{1808.05955}%
  \BibitemOpen
  \bibfield  {author} {\bibinfo {author} {\bibfnamefont {T.}~\bibnamefont
  {Brinckmann}}, \bibinfo {author} {\bibfnamefont {D.~C.}\ \bibnamefont
  {Hooper}}, \bibinfo {author} {\bibfnamefont {M.}~\bibnamefont
  {Archidiacono}}, \bibinfo {author} {\bibfnamefont {J.}~\bibnamefont
  {Lesgourgues}}, \ and\ \bibinfo {author} {\bibfnamefont {T.}~\bibnamefont
  {Sprenger}},\ }\href {\doibase 10.1088/1475-7516/2019/01/059} {\bibfield
  {journal} {\bibinfo  {journal} {JCAP}\ }\textbf {\bibinfo {volume} {01}},\
  \bibinfo {pages} {059} (\bibinfo {year} {2019})},\ \Eprint
  {http://arxiv.org/abs/1808.05955} {arXiv:1808.05955 [astro-ph.CO]}
  \BibitemShut {NoStop}%
\bibitem [{\citenamefont {Slosar}\ \emph {et~al.}(2019)\citenamefont {Slosar}
  \emph {et~al.}}]{1903.12016}%
  \BibitemOpen
  \bibfield  {author} {\bibinfo {author} {\bibfnamefont {A.}~\bibnamefont
  {Slosar}} \emph {et~al.},\ }\href@noop {} {\  (\bibinfo {year} {2019})},\
  \Eprint {http://arxiv.org/abs/1903.12016} {arXiv:1903.12016 [astro-ph.CO]}
  \BibitemShut {NoStop}%
\bibitem [{\citenamefont {Green}\ \emph {et~al.}(2019)\citenamefont {Green}
  \emph {et~al.}}]{1903.04763}%
  \BibitemOpen
  \bibfield  {author} {\bibinfo {author} {\bibfnamefont {D.}~\bibnamefont
  {Green}} \emph {et~al.},\ }\href@noop {} {\bibfield  {journal} {\bibinfo
  {journal} {Bull. Am. Astron. Soc.}\ }\textbf {\bibinfo {volume} {51}},\
  \bibinfo {pages} {159} (\bibinfo {year} {2019})},\ \Eprint
  {http://arxiv.org/abs/1903.04763} {arXiv:1903.04763 [astro-ph.CO]}
  \BibitemShut {NoStop}%
\bibitem [{\citenamefont {Bagla}(2005)}]{astro-ph/0411043}%
  \BibitemOpen
  \bibfield  {author} {\bibinfo {author} {\bibfnamefont {J.~S.}\ \bibnamefont
  {Bagla}},\ }\href@noop {} {\bibfield  {journal} {\bibinfo  {journal} {Curr.
  Sci.}\ }\textbf {\bibinfo {volume} {88}},\ \bibinfo {pages} {1088} (\bibinfo
  {year} {2005})},\ \Eprint {http://arxiv.org/abs/astro-ph/0411043}
  {arXiv:astro-ph/0411043} \BibitemShut {NoStop}%
\bibitem [{\citenamefont {Springel}(2005)}]{astro-ph/0505010}%
  \BibitemOpen
  \bibfield  {author} {\bibinfo {author} {\bibfnamefont {V.}~\bibnamefont
  {Springel}},\ }\href {\doibase 10.1111/j.1365-2966.2005.09655.x} {\bibfield
  {journal} {\bibinfo  {journal} {Mon. Not. Roy. Astron. Soc.}\ }\textbf
  {\bibinfo {volume} {364}},\ \bibinfo {pages} {1105} (\bibinfo {year}
  {2005})},\ \Eprint {http://arxiv.org/abs/astro-ph/0505010}
  {arXiv:astro-ph/0505010} \BibitemShut {NoStop}%
\bibitem [{\citenamefont {Lewis}\ and\ \citenamefont
  {Challinor}(2011)}]{lewis2011camb}%
  \BibitemOpen
  \bibfield  {author} {\bibinfo {author} {\bibfnamefont {A.}~\bibnamefont
  {Lewis}}\ and\ \bibinfo {author} {\bibfnamefont {A.}~\bibnamefont
  {Challinor}},\ }\href@noop {} {\bibfield  {journal} {\bibinfo  {journal}
  {ascl}\ ,\ \bibinfo {pages} {ascl}} (\bibinfo {year} {2011})}\BibitemShut
  {NoStop}%
\bibitem [{\citenamefont {Lesgourgues}(2011)}]{1104.2932}%
  \BibitemOpen
  \bibfield  {author} {\bibinfo {author} {\bibfnamefont {J.}~\bibnamefont
  {Lesgourgues}},\ }\href@noop {} {\  (\bibinfo {year} {2011})},\ \Eprint
  {http://arxiv.org/abs/1104.2932} {arXiv:1104.2932 [astro-ph.IM]} \BibitemShut
  {NoStop}%
\bibitem [{\citenamefont {Ma}\ and\ \citenamefont
  {Bertschinger}(1995)}]{astro-ph/9506072}%
  \BibitemOpen
  \bibfield  {author} {\bibinfo {author} {\bibfnamefont {C.-P.}\ \bibnamefont
  {Ma}}\ and\ \bibinfo {author} {\bibfnamefont {E.}~\bibnamefont
  {Bertschinger}},\ }\href {\doibase 10.1086/176550} {\bibfield  {journal}
  {\bibinfo  {journal} {Astrophys. J.}\ }\textbf {\bibinfo {volume} {455}},\
  \bibinfo {pages} {7} (\bibinfo {year} {1995})},\ \Eprint
  {http://arxiv.org/abs/astro-ph/9506072} {arXiv:astro-ph/9506072} \BibitemShut
  {NoStop}%
\bibitem [{\citenamefont {Lesgourgues}\ and\ \citenamefont
  {Tram}(2011)}]{1104.2935}%
  \BibitemOpen
  \bibfield  {author} {\bibinfo {author} {\bibfnamefont {J.}~\bibnamefont
  {Lesgourgues}}\ and\ \bibinfo {author} {\bibfnamefont {T.}~\bibnamefont
  {Tram}},\ }\href {\doibase 10.1088/1475-7516/2011/09/032} {\bibfield
  {journal} {\bibinfo  {journal} {JCAP}\ }\textbf {\bibinfo {volume} {09}},\
  \bibinfo {pages} {032} (\bibinfo {year} {2011})},\ \Eprint
  {http://arxiv.org/abs/1104.2935} {arXiv:1104.2935 [astro-ph.CO]} \BibitemShut
  {NoStop}%
\bibitem [{\citenamefont {Hu}(1998)}]{astro-ph/9801234}%
  \BibitemOpen
  \bibfield  {author} {\bibinfo {author} {\bibfnamefont {W.}~\bibnamefont
  {Hu}},\ }\href {\doibase 10.1086/306274} {\bibfield  {journal} {\bibinfo
  {journal} {Astrophys. J.}\ }\textbf {\bibinfo {volume} {506}},\ \bibinfo
  {pages} {485} (\bibinfo {year} {1998})},\ \Eprint
  {http://arxiv.org/abs/astro-ph/9801234} {arXiv:astro-ph/9801234} \BibitemShut
  {NoStop}%
\bibitem [{\citenamefont {Lewis}\ and\ \citenamefont
  {Challinor}(2002)}]{astro-ph/0203507}%
  \BibitemOpen
  \bibfield  {author} {\bibinfo {author} {\bibfnamefont {A.}~\bibnamefont
  {Lewis}}\ and\ \bibinfo {author} {\bibfnamefont {A.}~\bibnamefont
  {Challinor}},\ }\href {\doibase 10.1103/PhysRevD.66.023531} {\bibfield
  {journal} {\bibinfo  {journal} {Phys. Rev. D}\ }\textbf {\bibinfo {volume}
  {66}},\ \bibinfo {pages} {023531} (\bibinfo {year} {2002})},\ \Eprint
  {http://arxiv.org/abs/astro-ph/0203507} {arXiv:astro-ph/0203507} \BibitemShut
  {NoStop}%
\bibitem [{\citenamefont {Bertschinger}(2006)}]{astro-ph/0607319}%
  \BibitemOpen
  \bibfield  {author} {\bibinfo {author} {\bibfnamefont {E.}~\bibnamefont
  {Bertschinger}},\ }\href {\doibase 10.1103/PhysRevD.74.063509} {\bibfield
  {journal} {\bibinfo  {journal} {Phys. Rev. D}\ }\textbf {\bibinfo {volume}
  {74}},\ \bibinfo {pages} {063509} (\bibinfo {year} {2006})},\ \Eprint
  {http://arxiv.org/abs/astro-ph/0607319} {arXiv:astro-ph/0607319} \BibitemShut
  {NoStop}%
\bibitem [{\citenamefont {Howlett}\ \emph {et~al.}(2012)\citenamefont
  {Howlett}, \citenamefont {Lewis}, \citenamefont {Hall},\ and\ \citenamefont
  {Challinor}}]{1201.3654}%
  \BibitemOpen
  \bibfield  {author} {\bibinfo {author} {\bibfnamefont {C.}~\bibnamefont
  {Howlett}}, \bibinfo {author} {\bibfnamefont {A.}~\bibnamefont {Lewis}},
  \bibinfo {author} {\bibfnamefont {A.}~\bibnamefont {Hall}}, \ and\ \bibinfo
  {author} {\bibfnamefont {A.}~\bibnamefont {Challinor}},\ }\href {\doibase
  10.1088/1475-7516/2012/04/027} {\bibfield  {journal} {\bibinfo  {journal}
  {JCAP}\ }\textbf {\bibinfo {volume} {04}},\ \bibinfo {pages} {027} (\bibinfo
  {year} {2012})},\ \Eprint {http://arxiv.org/abs/1201.3654} {arXiv:1201.3654
  [astro-ph.CO]} \BibitemShut {NoStop}%
\bibitem [{\citenamefont {Dakin}\ \emph {et~al.}(2019)\citenamefont {Dakin},
  \citenamefont {Brandbyge}, \citenamefont {Hannestad}, \citenamefont
  {Haugb\o{}lle},\ and\ \citenamefont {Tram}}]{1712.03944}%
  \BibitemOpen
  \bibfield  {author} {\bibinfo {author} {\bibfnamefont {J.}~\bibnamefont
  {Dakin}}, \bibinfo {author} {\bibfnamefont {J.}~\bibnamefont {Brandbyge}},
  \bibinfo {author} {\bibfnamefont {S.}~\bibnamefont {Hannestad}}, \bibinfo
  {author} {\bibfnamefont {T.}~\bibnamefont {Haugb\o{}lle}}, \ and\ \bibinfo
  {author} {\bibfnamefont {T.}~\bibnamefont {Tram}},\ }\href {\doibase
  10.1088/1475-7516/2019/02/052} {\bibfield  {journal} {\bibinfo  {journal}
  {JCAP}\ }\textbf {\bibinfo {volume} {02}},\ \bibinfo {pages} {052} (\bibinfo
  {year} {2019})},\ \Eprint {http://arxiv.org/abs/1712.03944} {arXiv:1712.03944
  [astro-ph.CO]} \BibitemShut {NoStop}%
\bibitem [{\citenamefont {Shoji}\ and\ \citenamefont
  {Komatsu}(2010)}]{1003.0942}%
  \BibitemOpen
  \bibfield  {author} {\bibinfo {author} {\bibfnamefont {M.}~\bibnamefont
  {Shoji}}\ and\ \bibinfo {author} {\bibfnamefont {E.}~\bibnamefont
  {Komatsu}},\ }\href {\doibase 10.1103/PhysRevD.81.123516} {\bibfield
  {journal} {\bibinfo  {journal} {Phys. Rev. D}\ }\textbf {\bibinfo {volume}
  {81}},\ \bibinfo {pages} {123516} (\bibinfo {year} {2010})},\ \bibinfo {note}
  {[Erratum: Phys.Rev.D 82, 089901 (2010)]},\ \Eprint
  {http://arxiv.org/abs/1003.0942} {arXiv:1003.0942 [astro-ph.CO]} \BibitemShut
  {NoStop}%
\bibitem [{\citenamefont {Archidiacono}\ and\ \citenamefont
  {Hannestad}(2016)}]{1510.02907}%
  \BibitemOpen
  \bibfield  {author} {\bibinfo {author} {\bibfnamefont {M.}~\bibnamefont
  {Archidiacono}}\ and\ \bibinfo {author} {\bibfnamefont {S.}~\bibnamefont
  {Hannestad}},\ }\href {\doibase 10.1088/1475-7516/2016/06/018} {\bibfield
  {journal} {\bibinfo  {journal} {JCAP}\ }\textbf {\bibinfo {volume} {06}},\
  \bibinfo {pages} {018} (\bibinfo {year} {2016})},\ \Eprint
  {http://arxiv.org/abs/1510.02907} {arXiv:1510.02907 [astro-ph.CO]}
  \BibitemShut {NoStop}%
\end{thebibliography}%

\end{document}